\documentclass[10pt,journal,compsoc]{IEEEtran}

\usepackage{url}
\usepackage{graphicx}
\usepackage{subfigure}
\usepackage{color}
\usepackage{epsfig}
\usepackage{amssymb}
\usepackage{latexsym}
\usepackage{relsize}
\usepackage{amsmath,url}

\allowdisplaybreaks

\newcommand {\C} {{\rm I\kern-5.5pt C}}

\newcommand{\bP}[1]{{\mathbb{P}}\left[{#1}\right]}
\newcommand{\bE}[1]{{\mathbb{E}}\left[{#1}\right]}

\newcommand{\1}[1]{{\bf 1}\left[#1\right]}       % indicator 1{...}

\newcommand{\fsquare}{\vrule height6pt width7pt depth1pt}   % filled square
\newcommand{\myproof}{{\hfill \\ \bf Proof. \ }}           % Proof
\newcommand{\myendpf}{\hfill\fsquare \\[0.1in]}             % end of proof

\newtheorem{theorem}{Theorem}[section]

\newtheorem{lemma}[theorem]{Lemma}
\newtheorem{proposition}[theorem]{Proposition}
\newtheorem{corollary}[theorem]{Corollary}

\begin{document}

\title{Counting triangles, tunable clustering
and the small-world property in random key graphs \\
(Extended version)
\thanks{This work was supported in part by NSF Grants CCF-0729093 and
CCF-1617934. Part of the work was conducted during Fall 2014 
while A.M. Makowski was a Visiting Professor with the Department of Statistics
of the Hebrew University  of Jerusalem with the support of a fellowship
from the Lady Davis Trust. The authors also
thank a colleague (who wishes to remain anonymous) for suggestions that lead to a
shorter proof of the one law in Theorem \ref{thm:OneLaw}.}
\thanks{
Parts of the material were presented in 
the 47th Annual
Allerton Conference on Communication, Control and Computing,
Monticello (IL), September 2009, and 
in the First Workshop on
Applications of Graph Theory in Wireless Ad hoc Networks and
Sensor Networks (GRAPH-HOC 2009), Chennai (India), December 2009.
}
 \thanks{O. Ya\u{g}an 
 is with the Department of Electrical and Computer Engineering and 
CyLab, Carnegie Mellon University, 
Pittsburgh, PA 15213 USA (e-mail: oyagan@ece.cmu.edu).} 
\thanks{A. M. Makowski is with the
Department of Electrical and Computer Engineering, and the
Institute for Systems Research, University of Maryland, College
Park, MD 20742 USA (e-mail: armand@isr.umd.edu).}
}

\author{
Osman Ya\u{g}an,  {\it Member, IEEE} and Armand M. Makowski, \it{Fellow, IEEE}%\\
%{\tt osmanyagan@gmail.com}, {\tt armand@isr.umd.edu} \\
%Department of Electrical and Computer Engineering\\
%and Institute for Systems Research \\
%University of Maryland, College Park, MD 20742.\\
}

%\author{
%\IEEEauthorblockN{Osman Ya\u{g}an}
%\IEEEauthorblockA{Department of Electrical and Computer \\ Engineering,
%and CyLab \\
%Carnegie Mellon University,
%Moffett Field, CA 94035.\\
%Email: oyagan@ece.cmu.edu}
%\and
%\IEEEauthorblockN{Armand M. Makowski}
%\IEEEauthorblockA{Department of Electrical and  Computer \\
%                  Engineering,
%                  and Institute for Systems Research \\
%                  University of Maryland, College Park, MD 20742.\\
%Email: armand@isr.umd.edu}
%}
%
%

\IEEEtitleabstractindextext{
\begin{abstract}
%\normalsize 
Random key graphs were introduced  to study various properties of the Eschenauer-Gligor
key predistribution scheme for wireless sensor networks (WSNs). 
Recently this class of random graphs  has received much attention in contexts as diverse as
recommender systems, social network modeling, and clustering and classification analysis. 
This paper is devoted to analyzing various properties of random key graphs.
In particular, we establish a zero-one law for the
the existence of triangles in random key graphs, and identify the corresponding critical scaling. 
This zero-one law exhibits significant differences with the corresponding result in Erd\H{o}s-R\'enyi (ER) graphs.
We also compute the clustering coefficient of random key graphs, and
compare it to  that of ER graphs in the many node regime
when their expected average degrees are asymptotically equivalent.
For the parameter range of practical relevance in both wireless sensor network 
and social network applications, random key graphs are shown to be much more clustered than the corresponding
ER graphs. We also explore the suitability of random key graphs as small world models in the sense of Watts and Strogatz.
\end{abstract}

}

\maketitle
\IEEEdisplaynontitleabstractindextext
\IEEEpeerreviewmaketitle

\IEEEraisesectionheading{\section{Introduction}
\label{sec:Introduction}}

Random key graphs are random graphs that belong to the class of
random intersection graphs \cite{SingerThesis}; they are
also called uniform random intersection graphs by some
authors \cite{BlackburnGerke, GodehardtJaworski,
GodehardtJaworskiRybarczyk}. They have appeared recently in
application areas as diverse as epidemics in social networks \cite{BallSirlTrapman},
 clustering analysis
\cite{GodehardtJaworski, GodehardtJaworskiRybarczyk},
collaborative filtering in recommender systems \cite{Marbach2008},
and random key predistribution for wireless sensor networks (WSNs)
\cite{EschenauerGligor}. In this last context, random key graphs
naturally occur in the study of a random key
predistribution scheme introduced by Eschenauer and Gligor
\cite{EschenauerGligor}: Before deployment, each sensor in a WSN
is independently assigned $K$ distinct cryptographic keys which
are selected at random from a  large pool of $P$ keys. These
$K$ keys constitute the key ring of the sensor node and are inserted into
its memory module. Two sensor nodes can then establish a secure
edge between them if they are within transmission range of each
other and if their key rings have at least one key in common; see
\cite{EschenauerGligor} for implementation details. If we assume
{\em full visibility}, namely that nodes are all within communication
range of each other, then secure communication between two nodes
requires only that their key rings share at least one key. The
resulting notion of adjacency defines the class of random key
graphs; see Section \ref{sec:ModelDefinitions}
for precise definitions.

Much efforts have recently been devoted to developing zero-one
laws for the property of connectivity in random key graphs. A key
motivation can be found in the need to obtain conditions under
which the scheme of Eschenauer and Gligor guarantees secure
connectivity with high probability in large networks \cite{ICC2011}. An
interesting feature of this work lies in  the following fact:
Although random key graphs are {\em not} stochastically equivalent
to the classical Erd\H{o}s-R\'enyi graphs \cite{ER1960}, it is
possible to {\em formally} transfer well-known zero-one laws for connectivity in
Erd\H{o}s-R\'enyi graphs to random key graphs by asymptotically
matching their edge probabilities. This approach, which was
initiated by Eschenauer and Gligor in their original analysis
\cite{EschenauerGligor}, has now been validated rigorously; see
the papers \cite{BlackburnGerke,
DiPietroManciniMeiPanconesiRadhakrishnan2008,Rybarczyk2011,
YaganMakowskiISIT2009, YaganMakowskiConnectivity,ZhaoYaganGligor} for recent
developments. Rybarczyk \cite{Rybarczyk2011} has
shown that this transfer from Erd\H{o}s-R\'enyi graphs also works
for a number of issues related to the giant component and its
diameter.

In view of these developments, it is natural to wonder whether
this (formal) transfer technique applies to other graph properties. In
particular, in the literature on random graphs there is long
standing interest \cite{Bollobas, ER1960, JansonLuczakRucinski,
KaronskiScheinermanSingerCohen, PenroseBook, SingerThesis} in the
containment of certain (small) subgraphs, the simplest one being
the {\em triangle}. This particular case is also of some practical relevance: 
The number of triangles in a graph is closely related to its clustering coefficient,
and for random key graphs this has implications on network
resiliency under the EG scheme (e.g., see 
\cite{DiPietroManciniMeiPanconesiRadhakrishnan2008}) and also on its applicability and relevance in different domains including social networks -- more on that later.

With these in mind,
in the present paper we study the 
 triangle containment problem in random key graphs.
In particular, we establish a zero-one law for the existence of triangles and identify the
corresponding {\em critical} scaling.
By the help of this result (and its proof), we conclude that in the many node regime, the
expected number of triangles in random key graphs is always at
least as large as the corresponding quantity in asymptotically
matched Erd\H{o}s-R\'enyi graphs. For the parameter range of
practical relevance in WSNs, we show that 
this expected number of triangles can
be orders of magnitude larger in random key graphs than in
Erd\H{o}s-R\'enyi graphs, confirming the observations made earlier via
simulations by Di Pietro et al.
\cite{DiPietroManciniMeiPanconesiRadhakrishnan2008}. 

These results 
 show that
transferring results from Erd\H{o}s-R\'enyi graphs to random key graphs 
by matching
their edge probabilities is not a valid approach in general, and
can be quite misleading in the context of WSNs.
In particular, our results indicate that the asymptotic
equivalence of random key graphs and Erd\H{o}s-R\'enyi graphs (in the sense discussed
in \cite{SingerThesis}) is possible only when the size of key
rings is comparable to the network size, a case not very realistic
in WSNs due to the severe constraints imposed on the
memory and computational capabilities of sensors. 
This points to the inadequacy of Erd\H{o}s-R\'enyi
graphs to capture some key properties of the EG scheme in
realistic WSN implementations, and reinforces the call for a direct
investigation of random key graphs.

The number (and {\em fraction}) of triangles in a network is closely related to its
 {\em clustering coefficient}, a metric known to have
a significant impact on the dynamics of many interesting processes 
that take place on the network; e.g.,
the diffusion of information and epidemic diseases 
\cite{GleesonMelnikHackett, Miller,NewmanCluster,YongYaganInfoProp,info_prop_real_time}, 
the propagation of influence \cite{HackettMelnikGleeson,YongYaganInfluence}, and cascading failures \cite{HuangShaoStanley}.
With this in mind, we also study the clustering coefficient of random key graphs and compare it with that of an Erd\H{o}s-R\'enyi
graph. We observe that the clustering
coefficient of a random key graph is never smaller than the
clustering coefficient of the corresponding Erd\H{o}s-R\'enyi
graph with {\em identical} expected average degree. For the
parameter range that is relevant for large scale social networks (as well as WSNs), we
show that random key graphs are in fact {\em much more clustered} than
Erd\H{o}s-R\'enyi graphs when expected average degrees are
asymptotically equivalent. Recalling the fact that 
random key graphs also have a small {\em diameter}
\cite{Rybarczyk2011,YaganMakowskiCISS2010}, 
we then conclude that
random key graphs are {\em small-worlds} in the sense introduced by 
Watts and Strogatz \cite{Watts&Strogatz}. This reinforces the possibility of using random key graphs
in a wide range of applications including social network modeling.

%The zero-one law obtained here (for the existence of triangles) 
%were announced in the conference
%paper \cite{YaganMakowskiAllerton2009} with much longer proofs.
%The results regarding the clustering coefficient of random key
%graphs were presented in the conference paper
%\cite{YaganMakowskiChennai2009}. 

In line with results currently
available for other classes of graphs, e.g., Erd\H{o}s-R\'enyi
graphs \cite[Chap. 3]{JansonLuczakRucinski} and random geometric
graphs \cite[Chap. 3]{PenroseBook}, it would be interesting to
consider the containment problem for small subgraphs other than
triangles in the context of random key graphs. To the best of our knowledge, this issue has not been considered
in the literature.
 Future work may also consider 
other properties of random key graphs that might be relevant in various applications; e.g., Hamiltonicity, spectral radius, percolation, etc.

The paper is organized as follows:
We formally introduce the class of random key graphs in Section  \ref{sec:ModelDefinitions},
with various definitions for the clustering coefficient presented in Section \ref{sec:ClusteringDefns}.
In Section \ref{sec:CountingTriangles} we evaluate the first and second moments of
the number of triangles in random key graphs.
Our main results are presented in Section \ref{sec:MainResults}:
A zero-one law concerning the containment of triangles 
in random key graphs is discussed
in Section \ref{subsec:ZeroOneLawsExistenceTriangles}
while its clustering coefficient is computed in Section \ref{subsec:FixedParameters}.
%A number of useful asymptotic equivalences are presented in Section \ref{sec:AsymptoticEquivalences}.
Relevant definitions and facts concerning Erd\H{o}s-R\'enyi graphs are given in Section \ref{sec:FactsERs}.
Section \ref{sec:ComparingTriangles} and
Section \ref{sec:ComparingClusteringCoefficients}
are devoted to comparing random key graphs and Erd\H{o}s-R\'enyi graphs
in terms of their number of triangles and clustering coefficients,
respectively. 
Section \ref{sec:WSN_Implications}
and Section \ref{sec:SocialNetworks_Implications}
discuss the implications of our results on utilizing random key graph in
the context of WSN and social network applications, respectively.
The proofs of the main results of the paper are available in Section \ref{sec:proofs}, 
while some technical results are established in Section \ref{sec:proofs_prelim}.
% and in Appendix.
%Theorem \ref{thm:Equiv_of_clust_defn_RKG} and Theorem \ref{thm:Equiv_of_clust_defn_ER}
%are established in 
%Section \ref{sec:RKGConvergenceClustering}
%and Section \ref{sec:ERConvergenceClustering}, respectively.

A word on the notation and conventions in use: Unless specified otherwise,
all limiting statements, including asymptotic equivalences, are understood with
$n$ going to infinity. The random variables (rvs) under
consideration are all defined on the same probability triple
$(\Omega, {\cal F}, \mathbb{P})$; its construction is standard and omitted in
the interest of brevity. 
Probabilistic statements are made with respect to this probability measure $\mathbb{P}$, 
and we denote the corresponding expectation operator by $\mathbb{E}$. 
We denote almost sure convergence (under $\mathbb{P}$) by a.s.
The indicator function of an event $E$ is denoted by $\1{E}$. For any
discrete set $S$ we write $|S|$ for its cardinality. 
We denote almost sure convergence by a.s.

\section{Model and definitions}
\label{sec:ModelDefinitions}

\subsection{Random key graphs}
Pick positive integers $K$ and $P$ 
such that $K \leq P$, and fix $n=3, 4, \ldots $.
We shall group the integers $P$ and $K$ into the ordered pair 
$\theta \equiv (K,P)$ in order to lighten the notation. 

The model of interest here is parametrized by the number $n$ of nodes, the size $P$
of the key pool and the size $K$ of each key ring. For each
node $i=1, \ldots , n$, let $K_i(\theta)$ denote the random set of
$K$ distinct keys assigned to node $i$.
Thus, under the
convention that the $P$ keys are labeled $1, \ldots , P$, the
random set $K_i(\theta)$ is a subset of $\{ 1, \ldots , P \}$ with
$|K_i(\theta)| = K$. The rvs $K_1(\theta), \ldots , K_n(\theta)$
are assumed to be {\em i.i.d.}, each of which is {\em uniformly} 
distributed with
\begin{equation}
\bP{ K_i(\theta) = S } = {P \choose K} ^{-1}, \qquad i=1,\ldots, n
\label{eq:KeyDistrbution1}
\end{equation}
for any subset $S$ of $\{ 1, \ldots , P \}$ with $|S|=K$. This
corresponds to selecting keys randomly and {\em without}
replacement from the key pool.

Distinct nodes $i,j=1, \ldots , n$ are said to be adjacent if they
share at least one key in their key rings, namely
\begin{equation}
K_i (\theta) \cap K_j (\theta) \not = \emptyset ,
\label{eq:KeyGraphConnectivity}
\end{equation}
in which case an undirected edge is assigned between nodes $i$ and $j$. 
The adjacency constraints (\ref{eq:KeyGraphConnectivity})
define an undirected  random graph on the vertex set $\{ 1, \ldots , n\}$, 
hereafter denoted $\mathbb{K}(n; \theta )$. We refer to this random graph 
as the {\em random key graph}.

It is easy to check that
\begin{equation}
\bP{ K_i (\theta) \cap K_j (\theta) = \emptyset } = q(\theta)
\end{equation}
with
\begin{equation}
q (\theta) = \left \{
\begin{array}{ll}
0 & \mbox{if~ $P <2K$} \\
   & \\
\frac{{P-K \choose K}}{{P \choose K}} & \mbox{if~ $2K \leq P$.}
\end{array}
\right . 
\label{eq:q_theta}
\end{equation}
The probability $p(\theta)$ 
of edge occurrence between any two nodes is
therefore given by
\begin{equation}
p(\theta)
= 1 - q(\theta).
\label{eq:p_theta}
\end{equation}
If $P<2K$ there exists an edge between any pair of
nodes, and $\mathbb{K}(n;\theta)$ coincides with the complete
graph  on the vertex set $\{ 1, \ldots , n\}$. While it is always
the case that $0 \leq q(\theta) < 1 $,  it is plain from
(\ref{eq:q_theta}) that
$q(\theta)> 0$ if and only if $2K \leq P$. 

The expression (\ref{eq:q_theta}) is a 
consequence of the general fact
\begin{equation}
\bP{ S \cap K_i(\theta) = \emptyset } 
= 
\frac{{P- |S| \choose K}}{{P \choose K}}, 
\quad i=1, \ldots ,n
\label{eq:Probab_key_ring_does_not_intersect_S}
\end{equation}
valid for any subset $S$ of $\{ 1, \ldots , P \}$ 
with $|S| \leq P-K$.

We close by introducing the events
\[
E_{ij}(\theta) = 
\left [
K_i (\theta)  \cap K_j(\theta) \neq \emptyset
\right ] ,
\quad i,j =1, \ldots , n
\]
whose indicator functions
\[
\xi_{ij}(\theta) = \1{ K_i (\theta)  \cap K_j(\theta) \neq \emptyset },
\quad i,j =1, \ldots , n
\]
are the edge rvs defining the random key graph $\mathbb{K}(n;\theta)$.
For each $i=1, \ldots , n$,
it is a simple matter to check with the help of
(\ref{eq:Probab_key_ring_does_not_intersect_S}) that the events
$
\{
E_{ij}(\theta), \ j \neq i, 
j=1, \ldots , n 
 \}
$
are mutually independent, or equivalently, that the rvs
$
 \{
\xi_{ij}(\theta), \
j \neq i,
j=1, \ldots , n 
 \}$
form a collection of i.i.d. rvs.

\subsection{Clustering coefficient}
\label{sec:ClusteringDefns}

Many networks encountered in practice exhibit 
high clustering (or transitivity) in that 
the neighbors of a node are likely to be neighbors to each other \cite{SerranoBoguna}
-- Your friends are likely to be friends!
Clustering properties are known to have
a significant impact on the dynamics of many interesting processes 
that take place on a network, e.g.,
the diffusion of information and epidemic diseases 
\cite{GleesonMelnikHackett, Miller,NewmanCluster,YongYaganInfoProp}, 
the propagation of influence \cite{HackettMelnikGleeson,YongYaganInfluence}, and cascading failures \cite{HuangShaoStanley}.
With this in mind we shall investigate clustering in random key graphs under various parameter regimes.

A formal definition of clustering is given next. 
Consider an undirected graph $G$ with no self-loops on the vertex set $V$.
For each $i$ in $V$, let $T_i(G)$ denote the number of 
distinct {\em triangles} in $G$ that contain vertex $i$. 
The {\em local} clustering coefficient of {\em node} $i$ is given by
\begin{equation}
C_i (G)
=
\left \{
\begin{array}{ll}
\frac{ T_i (G) }
     { \frac{1}{2} d_i(d_i-1) }
  & \mbox{if $d_i \geq 2$} \\
  &              \\
0 & \mbox{otherwise} \\
\end{array}
\right .
\label{eq:ClusteringCoefficient}
\end{equation}
where $d_i$ is the degree of node $i$ in $G$.

There are, however, several possible definitions 
for a graph-{\em wide} notion of clustering \cite{Newman}:
Inspired by (\ref{eq:ClusteringCoefficient}), it is natural to consider
the {\em average} of the local clustering coefficient $C_{\rm Avg}(G)$ over the graph $G$, 
i.e.,
\begin{equation}
C_{\rm Avg}(G)
= 
\frac{1}{|V^\prime|} \sum_{i \in V^\prime } C_i(G) 
\label{eq:DefAvg}
\end{equation}
where $V^\prime = \{ i \in V: \ d_i \geq 2 \}$.
This last quantity, while natural, is often replaced
by the {\em global} clustering coefficient defined
as the \lq\lq fraction of transitive triples" over the whole graph $G$, namely,
\begin{equation}
C^{\star}(G) = 
\frac{ \sum_{i \in V} T_i (G) }
     { \frac{1}{2} \sum_{i \in V} d_i(d_i-1) }
\label{eq:Def2}
\end{equation}
provided $ \sum_{i \in V} d_i(d_i-1) > 0$. It is convenient to set
$ C^{\star}(G) = 0$ otherwise.

In the context of random graphs,
related (but simpler) definitions are possible when
the edge assignment rvs  are {\em exchangeable} 
(as is the case for the random graphs of interest here).
Recall that an undirected random graph $\mathbb{G}$ defined over the set of nodes $\{1, \ldots , n \}$
is characterized by the $\{0,1\}$-valued edge rvs
$\{ \xi_{ij} , \ i,j =1, \ldots , n \}$ with the interpretation that $\xi_{ij} = 1$ (resp. $\xi_{ij} = 0$) if
there is an edge (resp. no edge) between nodes $i$ and $j$.
As we consider graphs which are undirected with no self-loops, we impose the conditions
\[
\xi_{ij} = \xi_{ji}
\quad \mbox{and} \quad \xi_{ii} = 0,
\quad i,j=1, \ldots , n.
\]

A case of great interest arises when the rvs
$
\{ \xi_{ij} , \ 1 \leq i < j \leq n \}
$
form a family of {\em exchangeable} rvs \cite{AldousExchange}. In that setting,
a popular approach (e.g., see \cite{DeijfenKets}) is to define the clustering coefficient of the random graph 
$\mathbb{G}$ as the conditional probability
\begin{equation}
C( \mathbb{G}) 
= 
\bP{ E_{12} \;| \; E_{13} \cap E_{23} } 
\label{eq:ClustInRandomGraphs}
\end{equation}
where we have used the notation
\[
E_{ij} = \left [ \xi_{ij} = 1 \right ],
\quad  i,j=1, \ldots , n.
\]
For the random graphs considered here,
we show that the quantity (\ref{eq:ClustInRandomGraphs}) provides a good
approximation to the global clustering coefficient defined at (\ref{eq:Def2}), when $n$ is large; % and under appropriate scaling conditions;
see Theorem \ref{thm:Equiv_of_clust_defn_RKG} and Theorem \ref{thm:Equiv_of_clust_defn_ER}.
It is for this reason that we use the simpler definition (\ref{eq:ClustInRandomGraphs})
for studying clustering in the remainder of this paper.

\subsection{Counting triangles}
\label{sec:CountingTriangles} 

Pick positive integers $K$ and $P$ 
such that $K \leq P$, and fix $n=3, 4, \ldots $ 
For distinct $i, j, k= 1, \ldots, n$, we define the indicator function
\begin{align}
\chi_{{ijk}}(\theta) 
  = \1{ 
\begin{array}{c}
\mbox{Nodes $i$, $j$ and $k$ form} \\
\mbox{a~triangle~in~$\mathbb{K}(n; \theta)$} \\
\end{array}
}.
\label{eq:IndicatorTriangle}
\end{align}
The number of distinct triangles in $\mathbb{K}(n; \theta)$ 
is then simply given by
\begin{equation}
T_n (\theta) = {\sum}^n_{1 \leq i < j<k \leq n} \chi_{{ijk}}(\theta).
\label{eq:NumberOfTriangles}
\end{equation}
Of particular interest is the event that there exists at
least one triangle in $\mathbb{K}(n; \theta )$, namely
$[ T_n (\theta) >  0 ] = [ T_n (\theta) = 0 ]^{c}$.

One of our main results is a zero-one law for the existence of triangles in random key graphs. 
These results  will be established by the method of first and second moments
applied to the count variables (\ref{eq:NumberOfTriangles}),
e.g., see
\cite[p. 2]{Bollobas}, \cite[p. 55]{JansonLuczakRucinski}.
They are stated in terms of the quantity
\begin{equation}
\tau(\theta) = \frac{K^3}{P^2} + \left ( \frac{K^2}{P} \right)^3,
\quad
\begin{array}{c}
\theta = (K,P) \\
K,P =1,2, \ldots \\
\end{array}
\label{eq:Tau(Theta)}
\end{equation}
As we shall see soon in Proposition \ref{prop:AsymptoticEquivalence2}, this quantity gives the {\em asymptotic} probability of a triangle in random key graphs, when the parameters $K$ and $P$ are suitably scaled.
%most regimes of interest. 

Key to much of the discussion carried out in this paper are
the first two moments of the count variables
(\ref{eq:NumberOfTriangles}).
The first moment, computed next, 
will be conveniently expressed with the help
of the quantity $\beta (\theta)$ given by
\begin{equation}
\beta(\theta) = (1-q(\theta))^3 + q(\theta)^3 - q(\theta) r(\theta)
\label{eq:beta(theta)}
\end{equation}
with $r(\theta)$ defined by
\begin{equation}
r(\theta) 
= \left \{
\begin{array}{ll}
0 & \mbox{if~ $P <3K$} \\
& \\
\frac{{P-2K \choose K}}{{P \choose K}} & \mbox{if~ $3K \leq P$.}
\end{array}
\right . 
\label{eq:r(tetha)}
\end{equation}
Note that $r(\theta)$ corresponds to the probability
(\ref{eq:Probab_key_ring_does_not_intersect_S}) when $|S|=2K$.

\begin{proposition}
{\sl Fix $n=3,4, \ldots $. For positive integers $K$ and $P$ such
that $K \leq P$, we have
\begin{equation}
\bE{ \chi_{{123}} (\theta) } = \beta (\theta )
\label{eq:ProbTriangle}
\end{equation}
with $\beta(\theta)$ defined at (\ref{eq:beta(theta)}),
so that
\begin{equation}
\bE{ T_n (\theta) } = {n \choose 3 } \beta ( \theta ).
\label{eq:RKG+FirstMoment}
\end{equation}
} 
\label{prop:RKG+FirstMoment}
\end{proposition}
A proof of Proposition \ref{prop:RKG+FirstMoment} is given in
Section \ref{subsec:ProofRKG+FirstMoment}.  We see from (\ref{eq:ProbTriangle}) 
that
the quantity $\beta(\theta)$ gives the probability that three 
distinct vertices form a triangle in $\mathbb{K}(n;\theta)$.
For future reference, we note that
\begin{equation}
r(\theta) \leq q(\theta)^2
\label{eq:r(tetha)B}
\end{equation}
by direct inspection, whence
\begin{equation}
\beta (\theta) \geq (1 - q(\theta) )^3 > 0. 
\label{eq:r(tetha)C}
\end{equation}

The second moment of the count variables 
(\ref{eq:NumberOfTriangles}) is computed next; it 
will play a crucial role in the proofs of both
Theorem \ref{thm:OneLaw} and Theorem \ref{thm:Equiv_of_clust_defn_RKG}
that are forthcoming.

\begin{proposition}
{\sl For positive integers $K$ and $P$ such that $K \leq P$, we
have 
\begin{align}
\hspace{-1.3mm}\bE{ T_n (\theta)^2 }
 &=\bE{ T_n (\theta) }
 + \left (\hspace{-1mm}\frac{
{n-3\choose 3} }{{n \choose 3}} \hspace{-.5mm}+\hspace{-.5mm} 3 \frac{{n-3\choose 2}}{{n
\choose 3}}\hspace{-1mm} \right) 
 \left ( \bE{ T_n (\theta) } \right )^2
\nonumber \\
& ~~~~ + 3 (n-3) {n \choose 3}
 \cdot
\bE{\chi_{123}(\theta) \chi_{ 124} (\theta)}
\label{eq:SecondMoment}
\end{align}
for all $n=3,4, \ldots$
} 
\label{prop:SecondMoment}
\end{proposition}
The proof of Proposition \ref{prop:SecondMoment} is available in Section \ref{subsec:ProofRKG+SecondMoment}.

\section{Main results}
\label{sec:MainResults}

For simplicity of exposition we refer to any pair of functions
$P,K: \mathbb{N}_0 \rightarrow \mathbb{N}_0$ as a {\em scaling}
(for random key graphs)
provided the natural condition $K_n \leq P_n$ holds for all $n=3, 4, \ldots$. 

\subsection{Two asymptotic equivalences}

The two asymptotic equivalence results (under such scalings) presented next
will prove useful in a number of places. They provide easy asymptotic expressions
for the edge probability and for the probability of a triangle, respectively,
in large random key graphs.
The first one, already obtained in \cite{YaganMakowskiConnectivity}, 
is given here for easy reference.

\begin{lemma}
{\sl For any scaling $P,K: \mathbb{N}_0 \rightarrow \mathbb{N}_0$,
we have
\begin{equation}
\lim_{n \rightarrow \infty} q(\theta_n) = 1 
\quad \mbox{\sl if and only if} \quad 
\lim_{n \rightarrow \infty} \frac{K^2_n}{P_n} = 0,
\label{eq:Condition1+2}
\end{equation}
and under either condition at (\ref{eq:Condition1+2}),
the asymptotic equivalence
\begin{equation}
1 - q(\theta_n) \sim \frac{K^2_n}{P_n}
\label{eq:AsymptoticsEquivalence1}
\end{equation} 
holds.} 
\label{lem:AsymptoticEquivalence1}
\end{lemma}

The next result shows that under certain conditions the quantity
(\ref{eq:Tau(Theta)}) behaves asymptotically like 
 (\ref{eq:beta(theta)}) (which gives the probability that three nodes form a triangle in random key graphs). 
 
\begin{proposition}
{\sl For any scaling $P,K: \mathbb{N}_0 \rightarrow \mathbb{N}_0$
satisfying (\ref{eq:Condition1+2}), we have the asymptotic equivalence 
\begin{equation}
\beta(\theta_n) \sim \tau(\theta_n).
\label{eq:AsymptoticEquivalenceBeta=Tau}
\end{equation}
}
\label{prop:AsymptoticEquivalence2}
\end{proposition}
A proof of Proposition \ref{prop:AsymptoticEquivalence2} is given in Section \ref{sec:ProofOfAsymptEquiv2}.
In words, this result shows that under (\ref{eq:Condition1+2}) the probability of three 
vertices forming a triangle in random key graphs is asymptotically equivalent to 
\[
\tau(\theta_n) = \frac{K_n^3}{P_n^2} + \left ( \frac{K_n^2}{P_n} \right)^3.
\]

\subsection{Zero-one laws for the existence of triangles}
\label{subsec:ZeroOneLawsExistenceTriangles}

The zero-law, which is given first, is established in Section \ref{subsec:ProofZeroLaw}.

\begin{theorem}
{\sl For any scaling $P,K: \mathbb{N}_0 \rightarrow \mathbb{N}_0$,
the zero-law 
\begin{align*}
\lim_{n \rightarrow \infty }
\bP{T_n(\theta_n) > 0 } = 0
\end{align*}
holds under the condition
\begin{equation}
\lim_{n \rightarrow \infty } n^3 \tau(\theta_n) = 0 .
\label{eq:ConditionForZero}
\end{equation}
} 
\label{thm:ZeroLaw}
\end{theorem}

The one-law given next assumes a more involved form;
its proof is given in Section \ref{subsec:ProofOneLaw}.

\begin{theorem}
{\sl For any scaling $P,K: \mathbb{N}_0 \rightarrow \mathbb{N}_0$
for which the limit $\lim_{n \rightarrow \infty }
q(\theta_n) = q^\star $ exists, the one-law 
\[
\lim_{n \rightarrow \infty } \bP{T_n (\theta_n) > 0 } = 1
\] 
holds if either $ 0 \leq q^\star < 1$,
or if $q^\star =1$ and the additional condition
\begin{equation}
\lim_{n \rightarrow \infty } n^3 \tau(\theta_n) = \infty 
\label{eq:ConditionForOne}
\end{equation}
holds.
} 
\label{thm:OneLaw}
\end{theorem}

To facilitate an upcoming comparison with analogous results in
ER graphs, we combine Theorem \ref{thm:ZeroLaw} and
Theorem \ref{thm:OneLaw} into a single symmetric statement.

\begin{theorem}
{\sl For any scaling $P,K: \mathbb{N}_0 \rightarrow \mathbb{N}_0$
for which $\lim_{n \rightarrow \infty } q(\theta_n)$ exists, we
have
\begin{eqnarray}\nonumber
\lim_{n \rightarrow \infty } \bP{ T_n (\theta_n) > 0 }
 = \left \{
\begin{array}{ll}
0 & \mbox{if~ $\lim_{ n\rightarrow \infty } n^3  \tau (\theta_n) = 0 $} \\
& \\
1 & \mbox{if~ $\lim_{ n\rightarrow \infty } n^3  \tau (\theta_n) =
\infty $.}
\end{array}
\right . \label{eq:TriangleZeroOneLaw+RKG}
\end{eqnarray}
} 
\label{thm:TriangleZeroOneLaw+RKG}
\end{theorem}

By Lemma \ref{lem:AsymptoticEquivalence1}  
the condition $\lim_{ n\rightarrow \infty } n^3  \tau (\theta_n) = 0 $
implies $\lim_{n \rightarrow \infty } q(\theta_n) = 1$, hence
he limit $\lim_{n \rightarrow \infty } q(\theta_n)$ necessarily exists with
$q^\star = 1$.

\subsection{Clustering in random key graphs} 
\label{subsec:FixedParameters}

In accordance with definition (\ref{eq:ClustInRandomGraphs}), the clustering coefficient of
the random key graph $\mathbb{K}(n;\theta)$ is defined by
\begin{equation}
C_{\rm K}(\theta)
= 
\bP{ E_{12}(\theta) \;| \; E_{13}(\theta) \cap E_{23}(\theta) } .
\label{eq:ClustInRKGs}
\end{equation}
A closed form expression for this quantity is given next.

\begin{proposition}
{\sl For positive integers $K ,P$ such that $K \leq P$, we have
\begin{equation}
C_{\rm K}(\theta)
= 
\frac{ \beta(\theta)}{ ( 1 - q(\theta) )^2 } 
\label{eq:C_K(theta)} 
\end{equation}
with $\beta(\theta)$ given by (\ref{eq:beta(theta)}).
}
\label{prop:C_n_theta} 
\end{proposition}

\myproof
The definitions of $C_{\rm K}(\theta)$ and
$\chi_{{123}} (\theta)$ yield
\begin{eqnarray}
C_{\rm K}(\theta) 
= 
\frac{\bP{ E_{12} ( \theta ) \cap E_{13} ( \theta )
           \cap E_{23} ( \theta ) } }
     {\bP { E_{13} ( \theta ) \cap E_{23} (\theta) } } 
= \frac{ \bE{ \chi_{{123}} (\theta) } }{(1-q(\theta))^2}
\label{eq:clust_RK_start}
\end{eqnarray}
since the events $E_{13} ( \theta )$ and $E_{23} (\theta)$
are independent, with 
\begin{align}
&\bP { E_{13} ( \theta ) \cap E_{23} (\theta) 
}  
\nonumber \\
&=
\bP{ K_1(\theta) \cap K_3(\theta) \neq \emptyset,
     K_2(\theta) \cap K_3(\theta) \neq \emptyset
}
\nonumber \\
&=
\left ( 1 - q(\theta) \right )^2
\end{align}
by virtue of (\ref{eq:Probab_key_ring_does_not_intersect_S})
(and comments following it).
The conclusion (\ref{eq:C_K(theta)})
is immediate upon substituting
(\ref{eq:ProbTriangle}) into (\ref{eq:clust_RK_start}).
\myendpf

For random key graphs there is strong consistency between 
the definitions (\ref{eq:Def2}) and (\ref{eq:ClustInRKGs}) of clustering coefficient.

\begin{theorem}
{\sl For positive integers $K ,P$ such that $K \leq P$, we have
\begin{equation}
\lim_{n \rightarrow \infty} 
C^{\star}(\mathbb{K}(n;\theta)) 
=
C_{\rm K}(\theta)
\quad a.s.
\label{eq:Equiv_of_clust_defn_RKG}
\end{equation}
}
\label{thm:Equiv_of_clust_defn_RKG}
\end{theorem}

A proof of Theorem \ref{thm:Equiv_of_clust_defn_RKG} 
is given in Section \ref{sec:RKGConvergenceClustering}. 
To the best of our knowledge, Theorem \ref{thm:Equiv_of_clust_defn_RKG} 
is the first rigorous result in the literature that shows that the  conditional probability definition
(\ref{eq:ClustInRandomGraphs}) of clustering coefficient converges asymptotically almost surely to the empirical clustering coefficient measure of (\ref{eq:Def2}). For instance, Deijfen and Kets  indicated \cite{DeijfenKets}, for another class of random graphs, that the two definitions should be closely related, but that a rigorous proof would need significant additional work. 

%This should be closely
%related to the limiting quotient of the number of triangles and the number of triples with at least two edges present, which is one of the empirical measures of clustering that occur in the literature; see e.g. Newman (2003). Establishing this connection rigorously however requires additional arguments.

Simulation results given in Table I 
illustrate the convergence (\ref{eq:Equiv_of_clust_defn_RKG})
for several realistic parameter values. The numerical values of 
$C_{\rm K}(\theta)$
are obtained directly from the expressions 
(\ref{eq:ClustInRKGs}). The quantity
$\widehat C_n^\star (\theta)$ stands for the clustering coefficient 
of $\mathbb{K}(n;\theta)$, calculated through
(\ref{eq:Def2}) and averaged over $1000$ realizations; the number
of nodes is set to $n=1000$ in all simulations. The data support
the validity of (\ref{eq:Equiv_of_clust_defn_RKG}), and confirm 
the claim that for large networks  the quantity (\ref{eq:Def2}) captures essentially
the same structural information as (\ref{eq:ClustInRKGs}).

\begin{table}[t]
\begin{center}
\begin{tabular}{|c|c||c|c|c|}
%\hline \multicolumn{6}{|c|}{Clustering coefficients for fixed $\theta$ and $p=1-q(\theta)$} \\
\hline $K$ & $P$ & $1-q(\theta)$ & $~~C_K(\theta)~~$  & $~\widehat C_n^\star(\theta)~$ \\
\hline \hline
 $4$ & $10^3$  & 0.0159 & $0.2590$  & $0.2587$    \\
 $8$ & $5 \times 10^3$    & $0.0127$ & $0.1348$   & $0.1349$  \\
 $16$& $2 \times 10^4$    & $0.0127$ & $0.0737$   & $0.0736$  \\
 $20$& $4 \times 10^4$    & $0.0100$ & $0.0590$   & $0.0590$    \\
 $24$& $ 10^5$   & $0.0057$ & $0.0469$ & $0.0468$  \\
 $32$& $ 10^5$   & $0.0102$ & $0.0408$ & $0.0408$  \\
 $40$& $ 5 \times 10^5$   & $0.0032$ & $0.0280$ & $0.0280$ \\
 $64$& $ 10^6$   & $0.0041$ & $0.0196$ &  $0.0196$   \\ \hline
 \end{tabular}\vspace{3mm}
\caption{Clustering coefficients with fixed $\theta$ for random key graphs}
\end{center}
\vspace{-.3cm}
\end{table}

\section{Facts concerning Erd\H{o}s-R\'enyi graphs}
\label{sec:FactsERs}

A little later in this paper,
we shall compare random key graphs to related
Erd\H{o}s-R\'enyi (ER) graphs \cite{ER1960}, but first some notation: 
For each $n=2,3, \ldots $ and each $p$ in $[0,1]$,
let $\mathbb{G}(n;p)$ denote the
ER graph on the vertex set $\{ 1, \ldots , n\}$ 
with edge probability $p$. 
The ER graph $\mathbb{G}(n;p)$ is characterized by the fact that
the $\frac{n(n-1)}{2}$ possible undirected edges between the $n$
nodes are independently assigned with probability $p$. Thus, if
in analogy with earlier notation, with distinct $i,j=1, \ldots , n$,
we denote by $E_{ij}(p)$ the event that there is an
(undirected) edge between nodes $i$ and $j$ in $\mathbb{G}(n;p)$, then
the events $\{ E_{ij}(p), \ 1 \leq i < j \leq n \}$ are mutually independent, each of probability $p$. 
For ease of exposition
it will always be understood that $E_{ij}(p) = E_{ji}(p)$ 
for distinct $i, j =1, \ldots , n$.

Random key graphs are {\em not} stochastically equivalent
to ER graphs even when their edge probabilities are
matched exactly: As graph-valued rvs, the random graphs
$\mathbb{G}(n;p)$ and $\mathbb{K}(n; \theta)$ have different
distributions even under the {\em exact matching} condition
\begin{equation}
p = 1 - q(\theta) = p(\theta).
\label{eq:ExactEquivalence}
\end{equation}
See \cite{YaganMakowskiAllerton2009} for a discussion of
(dis)similarities. Under (\ref{eq:ExactEquivalence}) the random graphs
$\mathbb{G}(n;p)$ and $\mathbb{K}(n; \theta)$ are said to be exactly matched.

In analogy with (\ref{eq:NumberOfTriangles}) let $T_n(p)$ denote
the number of distinct triangles in $\mathbb{G}(n; p)$.
Under the enforced independence,
we note that
\begin{equation}
\bE{ T_n(p) } = {n \choose 3} \tau ^ \star (p), \quad
%\begin{array}{c}
%0 \leq p \leq 1 \\
n=3,4,  \ldots %\\
%\end{array}
\label{eq:ER+FirstMoment}
\end{equation}
with
\[
\tau^\star (p) = p^3,
\quad 0 \leq p \leq 1.
\]

The edge assignment rvs being exchangeable in ER graphs,
we can again define the clustering coefficient in $\mathbb{G}(n;p)$ 
according to (\ref{eq:ClustInRandomGraphs}) by setting
\begin{equation}
C_{\rm ER}(p) 
= \bP{E_{12}(p) \;| \; E_{13}(p) \cap E_{23}(p) }.
\label{eq:ClustInERGs}
\end{equation}
By mutual independence of the edge rvs it follows that
\begin{equation}
C_{\rm ER}(p)
=
\frac{ 
\bP{ E_{12}(p) \cap E_{13}(p) \cap E_{23}(p) }
}{ \bP{ E_{13}(p) \cap E_{23}(p) } }
= p
\label{eq:C_ER(p)}.
\end{equation}
Here as well, strong consistency holds between the two notions of clustering (\ref{eq:Def2}) and (\ref{eq:ClustInRKGs}).

\begin{theorem}
{\sl For every $p$ in $(0,1)$, we have
\begin{equation}
\lim_{n \rightarrow \infty} 
C^{\star}(\mathbb{G}(n;p)) 
=
C_{\rm ER}(p)
\quad a.s.
\label{eq:Equiv_of_clust_defn_ER}
\end{equation}
}
\label{thm:Equiv_of_clust_defn_ER}
\end{theorem}
This result can be established by arguments similar to the ones provided 
in the proof of Proposition \ref{thm:Equiv_of_clust_defn_RKG};  see
Appendix \ref{sec:ERConvergenceClustering} for details.
Table II expands on Table I given earlier in that we now
compare the clustering coefficients of exactly matched random key graphs 
and ER graphs for the parameter values used in Table I. 
The quantities for random key graphs are as before.
The numerical values of $C_{\rm ER}(p)$
are obtained directly from the expressions (\ref{eq:ClustInERGs}).
Here $\widehat C_n^\star (p)$
stands for the clustering coefficient of $\mathbb{G}(n; p)$. It is calculated through
(\ref{eq:Def2}) and averaged over $1000$ realizations. The number
of nodes is still set to $n=1000$ in all simulations. Again the data support
the claim that for large networks  the definition (\ref{eq:Def2}) captures essentially
the same information as the quantity (\ref{eq:ClustInERGs}).

\begin{table}[t]
\begin{center}
\begin{tabular}{|c|c||c|c|c|c|}
%\hline \multicolumn{6}{|c|}{Clustering coefficients for fixed $\theta$ and $p=1-q(\theta)$} \\
\hline $K$ & $P$ & $~~C_K(\theta)~~$  & $~\widehat C_n^\star(\theta)~$  & $~~C_{ER}(p)~~$
& $~\widehat C_n ^ \star (p)$ \\
\hline \hline
 $4$ & $10^3$  & $0.2590$ & $0.2587$   & $0.0159$ & $0.0159$ \\
 $8$ & $5 \times 10^3$    & $0.1348$   & $0.1349$ & $0.0127$ & $0.0128$ \\
 $16$& $2 \times 10^4$    & $0.0737$   & $0.0736$ & $0.0127$ & $0.0128$  \\
 $20$& $4 \times 10^4$    & $0.0590$   & $0.0590$ & $0.0100$ & $0.0100$   \\
 $24$& $ 10^5$   & $0.0469$ & $0.0468$   & $0.0057$  & $0.0057$ \\
 $32$& $ 10^5$   & $0.0408$ & $0.0408$   & $0.0102$  & $0.0102$  \\
 $40$& $ 5 \times 10^5$   & $0.0280$ & $0.0280$   & $0.0032$  & $0.0031$  \\
 $64$& $ 10^6$   & $0.0196$ & $0.0196$   & $0.0041$  & $0.0041$  \\ \hline
 \end{tabular}\vspace{3mm}
\caption{
Clustering coefficients with fixed $\theta$ and
$p=1-q(\theta)$ computed via (\ref{eq:q_theta})
}
\end{center}
\vspace{-.7cm}
\end{table}

Any mapping $p: \mathbb{N}_0 \rightarrow [0,1]$ will be called a scaling for ER graphs.
In order to meaningfully compare the asymptotic regime of
random key graphs with that of ER graphs under their respective scalings,
we shall say that the scaling 
$p: \mathbb{N}_0 \rightarrow [0,1]$ (for ER graphs) is
{\em asymptotically matched} to the scaling 
$P,K: \mathbb{N}_0 \rightarrow \mathbb{N}_0$ 
(for random key graphs) if
\begin{equation}
p_n \sim p(\theta_n) = 1 - q(\theta_n) .
\label{eq:AsymptoticEquivalence}
\end{equation}
Sometimes, when  (\ref{eq:AsymptoticEquivalence}) holds, we shall also say that
the random graphs $\mathbb{G}(n;p_n)$ and $\mathbb{K}(n; \theta_n)$ are asymptotically matched.
Under condition (\ref{eq:Condition1+2}), by Lemma \ref{lem:AsymptoticEquivalence1}
the asymptotic matching condition (\ref{eq:AsymptoticEquivalence}) amounts to
\begin{equation}
p_n \sim \frac{K_n^2}{P_n}.
\label{eq:AsymptoticMatchingEquiv}
\end{equation}

Condition (\ref{eq:ExactEquivalence})
(resp. (\ref{eq:AsymptoticEquivalence}))
is equivalent to requiring that
the expected degrees in 
$\mathbb{K}(n;\theta)$ and $\mathbb{G}(n;p)$ 
(resp. 
$\mathbb{K}(n;\theta_n)$ and $\mathbb{G}(n;p_n)$)
coincide (resp. are asymptotically equivalent).

\section{Comparing the number of triangles in random key graphs and ER graphs}
\label{sec:ComparingTriangles}

Fix $p$ in $(0,1]$, and 
positive integers $K$ and $P$ such that $K \leq P$.
From (\ref{eq:RKG+FirstMoment}) and (\ref{eq:ER+FirstMoment})
it is plain that
\begin{equation}
\frac{ \bE{ T_n(\theta) } }{ \bE{ T_n(p) } }
= 
\frac{ \beta (\theta)}
     { \tau^\star(p)},
\quad n=3,4, \ldots
\label{eq:RatioFirstMomentsBasic}
\end{equation}

Under the {\em exact} matching condition (\ref{eq:ExactEquivalence}),
with $p(\theta)$ given by (\ref{eq:p_theta}),
this last expression yields
\[
\frac{ \bE{ T_n(\theta) } }{ \bE{ T_n(p(\theta)) } }
=
\frac{ \beta (\theta)}
     { \tau^\star(p(\theta))}
=
1 
+ 
\frac{ q(\theta)^2 - r(\theta) }{ (1-q(\theta))^3}
\cdot q(\theta)
%\label{eq:RatioFirstMomentsExact}
\]
for each $n=3,4, \ldots$, whence
\[
\bE{ T_n(p(\theta)) } \leq \bE{ T_n(\theta) },
\quad n=3,4, \ldots
\]
by virtue of (\ref{eq:r(tetha)B}).
Consequently, the expected number of triangles in a random key graph is always 
at least as large as the corresponding quantity in an
ER graph exactly matched to it. 
This was already suggested by Di Pietro et al. \cite{DiPietroManciniMeiPanconesiRadhakrishnan2008} 
with the help of limited simulations.

An analogous result is available when the scalings are only
{\em asymptotically} matched.

\begin{corollary}
{\sl
Consider a scaling $K,P: \mathbb{N}_0 \rightarrow \mathbb{N}_0$
satisfying (\ref{eq:Condition1+2}),
and a scaling $p: \mathbb{N}_0 \rightarrow [0,1]$.
Under the asymptotic matching condition 
(\ref{eq:AsymptoticEquivalence}), we have the equivalence
\begin{equation}
\frac{ \bE{ T_n(\theta_n) } }
     { \bE{ T_n(p_n) } }
\sim 1 + \frac{P_n}{K_n^3} .
\label{eq:FirstMomentAsymptoticRatio}
\end{equation}
}
\label{cor:FirstMomentAsymptoticRatio}
\end{corollary}

In other words, for large $n$ the expected
number of triangles in random key graphs is always at least as
large as the corresponding quantity in asymptotically matched
ER graphs -- In fact, if the ratio $P_n/K_n^3 $ is large,
the number of triangles in random key graphs 
can be several orders of magnitude larger than that of ER graphs.
In Sections \ref{sec:WSN_Implications} and \ref{sec:SocialNetworks_Implications}
this issue is explored in the context of
wireless sensor networks and social networks, respectively.

\myproof
Replacing $\theta$ by $\theta_n$ and $p$ by $p_n$ 
according to the given scalings
in the expression (\ref{eq:RatioFirstMomentsBasic}),
we get
\[
\frac{ \bE{ T_n(\theta_n) } }
     { \bE{ T_n(p_n) } }
=
\frac{ \beta (\theta_n)}
     { \tau^\star(p_n)},
\quad n=3,4, \ldots
%\label{eq:RatioFirstMoments}
\]
Under  (\ref{eq:Condition1+2}), 
Proposition \ref{prop:AsymptoticEquivalence2} yields
\begin{equation}
\frac{ \bE{ T_n(\theta_n) } }
     { \bE{ T_n(p_n) } }
\sim 
\frac{ \tau (\theta_n)}
     { \tau^\star(p_n)}
\label{eq:RatioFirstMomentsEquivalent}
\end{equation}
with
\[
\frac{\tau(\theta_n)}
     {\tau^\star(p_n)}
= \frac{1}{p_n^3} \cdot \left ( \frac{K^3_n}{P^2_n} \right )
+
\frac{1}{p_n^3} \cdot \left ( \frac{K^2_n}{P_n} \right )^3, 
\quad n=3,4, \ldots
\]
With the help of (\ref{eq:AsymptoticMatchingEquiv}), we conclude
\begin{equation}
\frac{\tau(\theta_n)}
     {\tau ^ \star (p_n)} 
\sim 
1 + \frac{P_n}{K_n^3}
\label{eq:AsymptoticRatio_osy}
\end{equation}
and the equivalence
(\ref{eq:FirstMomentAsymptoticRatio}) follows from
(\ref{eq:RatioFirstMomentsEquivalent}).
\myendpf

From (\ref{eq:AsymptoticRatio_osy}) it follows
that under the asymptotic matching condition 
(\ref{eq:AsymptoticEquivalence}) (together with (\ref{eq:Condition1+2})), 
triangles will start appearing {\em earlier} in the evolution of a random key graph 
as compared to an ER graph (asymptotically) matched to it.
It should also be clear from (\ref{eq:AsymptoticRatio_osy}) that the larger the quantity $P_n/K_n^3$, 
the more pronounced will such difference be.

We close this section by comparing Theorem
\ref{thm:TriangleZeroOneLaw+RKG} with its analog for
ER graphs. 
Fix $n=3,4, \ldots $ and $p$ in $[0,1]$.
Consider the event that there exists at least one triangle in
$\mathbb{G}(n; p)$, i.e., $[ T_n(p) > 0 ]$. The following zero-one
law for triangle containment in ER graphs 
is well known 
\cite[Chap. 4]{Bollobas}, \cite[Thm. 3.4, p. 56]{JansonLuczakRucinski}.

\begin{theorem}
{\sl For any scaling $p: \mathbb{N}_0 \rightarrow [0,1]$, we have
\begin{eqnarray}\nonumber
\lim_{n \rightarrow \infty } \bP{ T_n( p_n ) > 0 }
 = \left \{
\begin{array}{ll}
0 & \mbox{if~ $\lim_{ n\rightarrow \infty } n^3  \tau ^ \star (p_n) = 0 $} \\
& \\
1 & \mbox{if~ $\lim_{ n\rightarrow \infty } n^3  \tau ^ \star (p_n) = \infty $.}
\end{array}
\right. 
\label{eq:ERZeroOneLawTriangle}
\end{eqnarray}
}
\label{thm:ERZeroOneLawTriangle}
\end{theorem}

This result, which  is also  established by the method of first and second moments, 
is easily understood once we recall (\ref{eq:ER+FirstMoment}).
As we compare Theorem \ref{thm:TriangleZeroOneLaw+RKG}
with Theorem \ref{thm:ERZeroOneLawTriangle}, we note a direct analogy since
the terms  $\tau(\theta_n)$ and $\tau ^ \star (p_n)$ 
correspond to the (asymptotic) probability that three arbitrary nodes form a triangle
in random key graphs and ER graphs, respectively. 

\section{Comparing the clustering coefficients of random key graphs and ER graphs}
\label{sec:ComparingClusteringCoefficients}

Fix $p$ in $(0,1]$, and 
positive integers $K$ and $P$ such that $K \leq P$.
Combining (\ref{eq:C_K(theta)}) and (\ref{eq:C_ER(p)}) we get
\begin{equation}
\frac{ C_{K}(\theta) }{ C_{\rm ER} (p) }
=
\frac{ \beta (\theta) }{ p (1-q(\theta) )^2 } .
\label{eq:ComparingClusteringCoefficientsBasic}
\end{equation}
Under the exact matching condition (\ref{eq:ExactEquivalence}) we find
\begin{eqnarray}
\frac{ C_{K}(\theta) }{ C_{\rm ER} (p(\theta)) }
&=&
\frac{ \beta(\theta)}{ (1-q(\theta)))^3 }
\nonumber \\
&=&
1 
+ 
\frac{ q(\theta)^2 - r(\theta) }
     { ( 1 - q(\theta) )^3 } \cdot q(\theta)
\label{eq:ComparingClusteringCoefficients}
\end{eqnarray}
as we recall (\ref{eq:p_theta}).
Thus, 
\begin{equation}
C_{\rm ER}(p(\theta)) \leq C_{\rm K}(\theta)
\label{eq:Inequality}
\end{equation}
by virtue of (\ref{eq:r(tetha)B}) --
The clustering coefficient of a random key graph 
is at least as large as that of the ER graph exactly matched to it. 

Several conclusions can be extracted from these expressions:
Equality in (\ref{eq:Inequality}) holds only
when $P < 2K$, i.e., from (\ref{eq:q_theta}) we get 
\[
\frac{ C_{\rm K} ( \theta ) } { C_{\rm ER} ( p ( \theta ) ) } = 1 
~~\mbox{if}~ K \leq P < 2K
\]
since then $q(\theta) = r(\theta) = 0$.
If $2K  \leq P < 3K$, then $0 < q(\theta)  < 1$ but $r(\theta)=0$, whence
\[
\frac{ C_{K}(\theta) }{ C_{\rm ER} (p(\theta)) } 
=
1 
+ 
\left ( \frac{ q(\theta) }{ 1 - q(\theta)  } \right )^3
> 1.
\]

Understanding the case $3K \leq P$ is more challenging due to a lack of simple expressions.
Therefore, before dealing with the case of an arbitrary positive integer $K$, we first consider
a couple of special cases as a way to explore the relative ranges possibly exhibited by the
clustering coefficients.
For $K=1$ it is a simple matter to check from
(\ref{eq:ComparingClusteringCoefficients}) that
\begin{equation}
\frac{ C_{\rm K} ( 1,P ) } { C_{\rm ER} ( p ( \theta ) ) } 
=
P
\label{eq:RatioK=1}
\end{equation}
for each $P=2,3, \ldots$.
For $K=2$ uninteresting calculations show that
\[
\frac{ C_{\rm K} ( 2 , P ) } { C_{\rm ER} ( p( \theta ) ) } 
= \frac{P}{2} 
\cdot 
\frac {2 P ^ 3 - 4 P ^ 2 - P + 3}{( 2 P -3)^3} 
\] 
for each $P=6,7, \ldots$, whence
\begin{equation}
\frac{P}{8} 
<
\frac{ C_{\rm K} ( 2 ,P ) } { C_{\rm ER} ( p ( \theta ) ) } 
< P
\label{eq:RatioK=2}
\end{equation}
on that range.
This upper bound is seen to hold by noting that
$4(2P^3 -4P^2 -P + 3 )
= (2P-3)^3 + (P-1)(20P-38) + 1 > (2P-3)^3$
for all $P=2,3, \ldots $. The lower bound follows from the easily checked fact that
$ 2 P ^ 3 - 4 P ^ 2 - P + 3 < 2 ( 2 P -3)^3$ for all $P=2,3, \ldots $.

The cases $K=1$ and $K=2$ may not be interesting from the 
perspective of envisioned modeling applications
of random key graphs. However, the discussion already shows that the parameters of the corresponding random key graph
can be selected (e.g., by taking $P$ very large  in these two cases) so that it has a much larger clustering coefficient than the ER graph 
exactly matched to it.
Additional limited numerical evidence along these lines is also available
in Table II discussed earlier. In fact, for any {\em given} $K$ we see that the linear behavior
found in (\ref{eq:RatioK=1}) and (\ref{eq:RatioK=2}) holds asymptotically for large $P$.

\begin{corollary}
{\sl
For each positive integer $K$, it holds that
\begin{equation}
\frac{C_{\rm K}(\theta)}{C_{\rm ER}(p(\theta))} \sim 1 + \frac { P } { K^3 }
\quad (P \rightarrow \infty).
\label{eq:RatioClusteringAsymptoticFixedK}
\end{equation}
}
\label{cor:RatioClusteringAsymptoticFixedK}
\end{corollary}

Thus, exactly matched random key graphs and ER graphs
will have vastly different clustering coefficients when $P$ is large.
This will be especially so for WSNs where the size of the key pool $P$ in
the Eschenauer-Gligor scheme is expected to be in the
range $2^{17}-2^{20}$ (with $K$ much smaller) \cite{EschenauerGligor}. 

\myproof
Fix positive integers $K$ and $P$ such that $2K \leq P$.
We can rewrite (\ref{eq:ComparingClusteringCoefficients}) as
\begin{equation}
\frac{ C_{K}(\theta) }{ C_{\rm ER} (p(\theta)) }
=
1 + \left ( \frac{ q(\theta) }{ 1 - q(\theta) }  \right )^3 
\cdot
\left ( 1 - \frac{r(\theta) }{q(\theta)^2} \right ).
\label{eq:ComparingClusteringCoefficientsB}
\end{equation}

With $K$ fixed and $P$ getting large, we see from Lemma \ref{lem:AsymptoticEquivalence1}
that
$1-q(\theta) \sim \frac{K^2}{P}$ and $q(\theta) \sim 1$
($P \rightarrow \infty$), so that
\[
\left ( \frac{ q(\theta) }{ 1 - q(\theta) }  \right )^3 
\sim \left ( \frac{P}{K^2} \right )^3
\quad (P \rightarrow \infty).
\]
The arguments given 
in the proof of Proposition \ref{prop:AsymptoticEquivalence2} to establish 
(\ref{eq:AsymptoticsEquivalence2Reduced}) can also be used to establish
\begin{equation}
1 - \frac{r(\theta)}{q(\theta)^2} \sim \frac{K^3}{P^2} 
\quad (P \rightarrow \infty).
\label{eq:AsymptoticsEquivalence2ReducedFixedK}
\end{equation}
Collecting we conclude to the validity of (\ref{eq:RatioClusteringAsymptoticFixedK}).
\myendpf

Next we compare the clustering coefficients of asymptotically matched
random key graphs and ER graphs when the parameters $\theta$ and $p$ are scaled with $n$.

\begin{corollary}
{\sl
Consider a scaling $K,P: \mathbb{N}_0 \rightarrow \mathbb{N}_0$
satisfying (\ref{eq:Condition1+2}) 
and a scaling $p: \mathbb{N}_0 \rightarrow [0,1]$.
Under the asymptotic matching condition 
(\ref{eq:AsymptoticEquivalence}), we have the equivalence
\begin{equation}
\frac{C_{\rm K}(\theta_n)}{C_{\rm ER}(p_n)}
\sim 
1 + \frac { P_n } { K_n ^ 3 } .
\label{eq:RatioClusteringAsymptotic}
\end{equation}
}
\label{cor:RatioClusteringAsymptotic}
\end{corollary}

\noindent  {\bf Proof.} \ 
As we replace $\theta$ by $\theta_n$ and $p$ by $p_n$ 
according to these scalings
in the expression (\ref{eq:ComparingClusteringCoefficientsBasic}),
we get
\begin{equation}
\frac{C_{\rm K}(\theta_n)}{C_{\rm ER}(p_n)}
=
\frac{ \beta(\theta_n) }{ p_n (1-q(\theta_n) )^2 },
\quad n= 3,4, \ldots
\label{eq:ComparingClusteringCoefficients1}
\end{equation}
Note that
\begin{equation}
\frac{C_{\rm K}(\theta_n)}{C_{\rm ER}(p_n)}
\sim
\frac{ \beta(\theta_n) }{ (1-q(\theta_n) )^3 }
\sim
\frac{ \tau(\theta_n) }{ (1-q(\theta_n) )^3 }.
\label{eq:ComparingClusteringCoefficients2}
\end{equation}
The first equivalence is a consequence of
(\ref{eq:AsymptoticEquivalence}) 
while the  second equivalence follows by
Proposition \ref{prop:AsymptoticEquivalence2} under (\ref{eq:Condition1+2}).
With (\ref{eq:AsymptoticMatchingEquiv}) being still valid here,
we easily conclude (\ref{eq:RatioClusteringAsymptotic})
by the same arguments as the ones used
to obtain (\ref{eq:FirstMomentAsymptoticRatio}).
\myendpf

Under (\ref{eq:Condition1+2}) and (\ref{eq:AsymptoticEquivalence}), 
we conclude that
\begin{equation}
\lim_{n \rightarrow \infty } 
\frac{C_{\rm K}(\theta_n)}{C_{\rm ER}(p_n)} = 1 
\quad \mbox{if} \quad 
\lim_{n \to \infty} \frac{ K_n ^ 3 } { P_n } = \infty, 
\label{eq:ConditionAsymptoticIndependence}
\end{equation}
and
\begin{equation}
\lim_{n \rightarrow \infty } 
\frac{C_{\rm K}(\theta_n)}{C_{\rm ER}(p_n)} = \infty
\quad \mbox{if} \quad 
\lim_{n \to \infty} \frac{ K_n ^ 3 } { P_n } = 0. 
\label{eq:ConditionAsymptoticHigherClustering}
\end{equation}
Thus, asymptotically matched random key graphs and ER graphs
can in principle have vastly different clustering coefficients. 
We explore this possibility in the next two sections where the implications
of the main results are discussed in the context of wireless sensor networks and 
of social networks based on common interest relationships.

\section{Wireless Sensor Networks}
\label{sec:WSN_Implications}

Random key graphs were originally introduced to model the random key pre-distribution
scheme proposed by Eschenauer and Gligor \cite{EschenauerGligor} in the context of WSNs.
When the WSN comprises $n$ nodes, it is natural to select the parameters
$K_n$ and $P_n$ in order for the induced random key graph to be
{\em connected}. However, there is a tradeoff between connectivity and
security \cite{DiPietroManciniMeiPanconesiRadhakrishnan2008},
requiring that $\frac{K_n ^ 2}{P_n}$ be kept as close as possible to
the critical scaling $\frac{\log n}{n}$ for connectivity (but above it); see the
papers \cite{BlackburnGerke,
DiPietroManciniMeiPanconesiRadhakrishnan2008, Rybarczyk2011,
YaganMakowskiISIT2009, YaganMakowskiConnectivity}. 
The desired regime near the boundary can be achieved by taking
\begin{equation}
\frac{K_n ^ 2 }{ P_n } \sim c \cdot \frac{ \log n }{ n }
\label{eq:K^2/P_sim_logn/n}
\end{equation}
with $c > 1$ but close to one. 

Now, consider the situation where the random key graph $\mathbb{K}(n;\theta_n)$ is
matched asymptotically to the ER random graph $\mathbb{G}(n;p_n)$ under
the asymptotic matching condition (\ref{eq:AsymptoticEquivalence}).
It follows from (\ref{eq:FirstMomentAsymptoticRatio}) that
\begin{equation}
\frac{ \bE{ T_n(\theta_n) } }
       { \bE{ T_n(p_n) } }
\sim 1 \quad \mbox{if and only if} \quad \frac{P_n}{K_n^3} = o(1)
\label{eq:NotPractical+1}
\end{equation}
under the condition (\ref{eq:Condition1+2}).
This last condition obviously occurs when (\ref{eq:K^2/P_sim_logn/n}) holds, 
in which case the condition at (\ref{eq:NotPractical+1}) amounts to taking
\[
\frac{1}{K_n} =  o(1) \left ( c \cdot \frac{\log n}{n} \right ) .
\]
Thus, under the connectivity condition (\ref{eq:K^2/P_sim_logn/n})  it holds that
\begin{equation}
\frac{ \bE{ T_n(\theta_n) } }
     { \bE{ T_n(p_n) } }
\sim 1 \quad \mbox{if and only if} \quad 
\lim_{n \rightarrow \infty} \frac{K_n}{ n / \log n } = \infty.
\label{eq:NotPractical}
\end{equation}
The expected number of triangles in random key graphs
is then of the same order as the corresponding quantity in
asymptotically matched ER graphs with 
$\bE{T_n(\theta_n) } \sim \bE{ T_n(p_n) } \sim \frac{c^3}{6} \left ( \log n \right )^3$
--  This is a direct consequence of
(\ref{eq:ER+FirstMoment}),
(\ref{eq:AsymptoticMatchingEquiv}) and
(\ref{eq:K^2/P_sim_logn/n}). This conclusion holds regardless of
the value of $c$ in (\ref{eq:K^2/P_sim_logn/n}).

However, given the limited memory and computational power of the
sensor nodes, key ring sizes satisfying (\ref{eq:NotPractical}) are
not practical since requiring $K_n \gg \frac{n}{\log n}$.
Furthermore, they will also result in {\em high} node degrees,
and this in turn will decrease network {\em resiliency}
against node capture attacks. It was proposed by Di Pietro
et al. \cite[Thm. 5.3]{DiPietroManciniMeiPanconesiRadhakrishnan2008}
that resiliency in large WSNs against node capture attacks 
can be ensured by selecting $K_n$ and $P_n$ such that
$\frac{K_n}{P_n} \sim \frac{1}{n}$. 
Under (\ref{eq:K^2/P_sim_logn/n}) this additional requirement then leads
to $K_n \sim c \cdot \log n$, whence $P_n \sim c \cdot n \log n$,
and (\ref{eq:FirstMomentAsymptoticRatio}) now implies
\begin{equation}
\lim_{n \rightarrow \infty} 
\frac{ \bE{ T_n(\theta_n) } }
     { \bE{ T_n(p_n) } }
= \lim_{n \rightarrow \infty} \left ( 1 + \frac{n}{(c \cdot \log
n) ^2} \right ) = \infty. \label{eq:ratio_infty}
\end{equation}
Therefore, for such realistic WSN implementations the expected number of
triangles in the induced random key graphs will be orders of
magnitude larger than in ER graphs.

Concerning the clustering coefficients, we see that under the condition (\ref{eq:K^2/P_sim_logn/n}),
(\ref{eq:ConditionAsymptoticIndependence}) can hold
only if the key ring size $K$ is much larger than $n/\log n$. As
already discussed, this condition can not be satisfied in a
practical WSN scenario due to storage limitations at the sensor nodes and 
security constraints. In fact, we see from
(\ref{eq:FirstMomentAsymptoticRatio}) and (\ref{eq:ratio_infty})
that, in a realistic WSN, the condition
(\ref{eq:ConditionAsymptoticHigherClustering}) is always in effect
and the clustering coefficient of the random key graph is much
larger than that of the asymptotically matched ER graph. 
%This provides yet another property of the random key graphs
%that ER graph models cannot adequately capture, further indicating
%the non-equivalence of the two models for practical WSN implementations.

\section{Social networks -- Can random key graphs be small worlds?}
\label{sec:SocialNetworks_Implications}

With an obvious change in terminology, random key graphs can be used 
to model certain types of social networks, e.g., see \cite{BallSirlTrapman,ZhaoYaganGligor}:
Instead of viewing $\{1, \ldots , P \}$ as a collection of cryptographic keys randomly assigned to the nodes
of a WSN according to the Eschenauer-Gligor scheme,
we can think of it as a list of \lq\lq interests," e.g., hobbies, books, movies, sports, etc., which are pursued 
by the members of a social group.
In that reformulation, the i.i.d. random sets $K_1(\theta), \ldots , K_n(\theta)$ 
appearing in the definition of the random key graph $\mathbb{K}(n;\theta)$ can now be interpreted as the interests assigned 
to the individual members of that group.\footnote{Here we 
assume that each individual has exactly $K$ interests drawn from the list $\{1, \ldots, P \}$.
More realistic models can be obtained through more complex randomization mechanisms 
as in the work of Godehardt et al. on general random intersection graphs
\cite{GodehardtJaworski, GodehardtJaworskiRybarczyk} and as in the work of Ya\u{g}an   on {\em inhomogeneous} random key graphs \cite{YaganInhomegeneous}.
}
The random key graph $\mathbb{K}(n;\theta)$ then naturally describes 
a {\em common-interest} relationship between community members
since two individuals are now adjacent in $\mathbb{K}(n;\theta)$ when they have at least one interest in common.

In parallel to the discussion given in
Section \ref{sec:WSN_Implications} for WSNs, we explore the parameter ranges 
likely to appear in practice for random key graphs modeling social networks. 
We do so with an eye towards understanding the behavior of
the expression appearing at (\ref{eq:FirstMomentAsymptoticRatio}).

We begin with the observation that most real-world social networks are known to 
be {\em sparse} in the sense that the expected number of edges per node appears to
remain (nearly) constant as the size of the network increases.
In the case of random key graphs, the expected degree of a node is given by $(n-1) (1-q(\theta_n))$ 
and sparsity amounts to $1-q(\theta_n) \sim \frac{c}{n}$ for some $c>0$,
or equivalently, to
\begin{align}
\frac{K_n^2}{P_n} \sim \frac{c}{n}
\label{eq:Sparsity}
\end{align}
by virtue of Lemma \ref{lem:AsymptoticEquivalence1}, whence
\begin{align}
\frac{P_n}{K_n^3} \sim \frac{n}{c K_n}.
\label{eq:SparsityAgain}
\end{align}

In view of Corollary \ref{cor:FirstMomentAsymptoticRatio} and Corollary \ref{cor:RatioClusteringAsymptotic},
in the sparse regime, random key graphs will have many more triangles 
and will be much more clustered (by orders of magnitude) than the asymptotically matched ER graphs unless
\begin{align}
\limsup_{n \rightarrow \infty} \frac{n}{K_n} < \infty.
\label{eq:NotPractical_2a}
\end{align}
This condition is equivalent to
\begin{align}
K_n = \Omega (n),
\label{eq:NotPractical_2c}
\end{align}
and is even more stringent than the corresponding condition (\ref{eq:NotPractical}) derived for WSN applications.
More importantly, under the condition (\ref{eq:Sparsity}) we have
$P_n \sim c^{-1} n K^2_n$, requiring 
\[
P_n = \Omega (n^3)
\]
if (\ref{eq:NotPractical_2c}) is also enforced. 
Thus, under (\ref{eq:Sparsity}) and (\ref{eq:NotPractical_2c}) generating the random key graph will require
each of the $n$ nodes to choose $K_n = \Omega (n)$ objects from a universe of size $P_n = \Omega(n^3)$.
The computational complexity of this task quickly becomes prohibitively high 
as the number of individuals in the social network becomes large. 
Yet, we would expect the realistic values for the number $K_n$ of interests of a single individual 
to be much smaller than the network size $n$, in sharp contrast with (\ref{eq:NotPractical_2c}). 
In other words, the condition (\ref{eq:NotPractical_2c}) is naturally eliminated in realistic 
applications of random key graphs to such social networks  -- The resulting random key graphs
will naturally have very high clustering and contain very large number of triangles when used for social network modeling.

Since random key graphs can be highly clustered, a natural
question arises as to their suitability for modeling the {\em small world effect}. 
This notion is linked to a well-known series of
experiments conducted by Milgram \cite{Milgram} in the late sixties. 
The results, commonly known as six degrees of separation,
suggest that the social network of people in the United States is
{\em small} in the sense that path lengths between pairs of
individuals are short. As a way to capture Milgram's experiments,
Watts and Strogatz \cite {Watts&Strogatz} introduced {\em small world} 
network models that are highly clustered and yet have a
small average path length. More precisely, a random graph is
considered to be a small world if its average path length is of
the same order as that of an ER graph with the same
expected average degree, but with a much larger clustering
coefficient.

The results of this paper already show that random key graphs can
satisfy the high clustering coefficient requirement of a small
world.  Under (\ref{eq:K^2/P_sim_logn/n}),
Rybarczyk  \cite{Rybarczyk2011} has shown that
\[
\textrm{diam}[\mathcal{K}(n;\theta_n)] \sim \frac{\log n}{\log
\log n}
\]
with high probability where $\mathcal{K}(n;\theta_n)$ is the
largest connected component of $\mathbb{K}(n;\theta_n)$. 
This suggests that the diameter, hence the average path length, in
random key graphs is {\em small} as was the case with
ER graphs \cite{ChungLu2001}. We also note
\cite[Corollary 5.2]{YaganMakowskiCISS2010} that random key
graphs have {\em very small} (e.g., $\leq 2$) diameter under certain
parameter ranges (e.g., with $P_n= O( n^{\delta} )$ with $0 < \delta < \frac{1}{2}$). 
Thus, random key graphs may indeed be
considered good candidate models for small worlds!

\section{Proofs of the preliminary results}
\label{sec:proofs_prelim}

In Sections \ref{subsec:ProofRKG+FirstMoment} and  \ref{subsec:ProofRKG+SecondMoment},
we fix positive integers $K$ and $P$ such
that $K \leq P$, and $n=3, 4, \ldots $. 

\subsection{A proof of Proposition \ref{prop:RKG+FirstMoment}}
\label{subsec:ProofRKG+FirstMoment}

As exchangeability yields
(\ref{eq:RKG+FirstMoment}),
we need only show the validity of (\ref{eq:ProbTriangle}).
%In the discussion that follows  we omit the explicit dependence on
%$\theta$ when no confusion arises from doing so. Also, 
We make
repeated use of the fact that for any
pair of events $E$ and $F$ in ${\cal F}$, 
we have 
\begin{equation}
\bP{ E \cap F } 
= 
\bP{ E} - \bP{ E \cap F^c }. 
\label{eq:Union+Intersection}
\end{equation}
Thus, by repeated application of (\ref{eq:Union+Intersection}) we find
\begin{eqnarray}
\lefteqn{\bE{ \chi_{{123}} (\theta) }} &&
\nonumber \\
&=& \bP{ 
\begin{array}{c} 
K_1(\theta)  \cap K_2(\theta) \neq \emptyset ,
K_1(\theta) \cap K_3(\theta)  \neq \emptyset ,\\
K_2(\theta)  \cap K_3(\theta)  \neq \emptyset \\
\end{array} 
}
\nonumber \\
&=& \bP{ K_1(\theta)  \cap K_2(\theta) \neq \emptyset ,
     K_1(\theta)  \cap K_3 (\theta)  \neq \emptyset }
\nonumber \\
& & ~ - \bP{ 
\begin{array}{c}
K_1(\theta)  \cap K_2 (\theta)  \neq \emptyset , K_1(\theta) \cap K_3(\theta)  \neq \emptyset , \\
K_2(\theta)  \cap K_3(\theta)  = \emptyset \\
\end{array}
}
\nonumber \\
&=& \bP{ K_1(\theta) \cap K_2(\theta)  \neq \emptyset ,
     K_1(\theta) \cap K_3(\theta)  \neq \emptyset }
     \nonumber \\
     & & ~
- \bP{ K_1(\theta)  \cap K_2(\theta) \neq \emptyset ,
             K_2(\theta)  \cap K_3(\theta) = \emptyset }
\nonumber \\
& & ~ + \bP{ 
\begin{array}{c}
K_1(\theta) \cap K_2(\theta) \neq \emptyset , K_1(\theta) \cap K_3(\theta) = \emptyset , \\
K_2 (\theta) \cap K_3(\theta)  = \emptyset \\
\end{array}
} .
\nonumber
\end{eqnarray}
By independence, with the help of
(\ref{eq:Probab_key_ring_does_not_intersect_S}), we readily obtain
the expressions
\[
\bP{ K_1(\theta) \cap K_2(\theta) \neq \emptyset , K_1(\theta) \cap K_3(\theta) \neq \emptyset } =
\left ( 1 - q(\theta) \right )^2 
%\label{eq:exp1}
\]
and
\begin{eqnarray}
& &
\bP{ K_1(\theta) \cap K_2(\theta) \neq \emptyset , K_2(\theta)  \cap K_3(\theta) = \emptyset }
\nonumber \\
&=&
\left ( 1 - q(\theta) \right ) q(\theta) .
\nonumber 
\end{eqnarray}

Next, as we use (\ref{eq:Union+Intersection})
one more time, we get
\begin{eqnarray}
\lefteqn{\bP{ 
\begin{array}{c}
K_1(\theta) \cap K_2(\theta) \neq \emptyset ,
K_1(\theta) \cap K_3(\theta) = \emptyset , \\
K_2(\theta) \cap K_3(\theta)  = \emptyset \\
\end{array}
}
 } & &
\nonumber \\
&=& \bP{ K_1(\theta) \cap K_3(\theta) = \emptyset ,
K_2(\theta) \cap K_3(\theta) = \emptyset }
\nonumber \\
& & ~  - \bP{ 
\begin{array}{c}
K_1(\theta) \cap K_2(\theta) = \emptyset , K_1(\theta) \cap K_3(\theta) = \emptyset \\
K_2(\theta) \cap K_3(\theta)  = \emptyset  \\
\end{array}
}.
\nonumber
\end{eqnarray}
Again, by independence, with the help of
(\ref{eq:Probab_key_ring_does_not_intersect_S}) we conclude that
\[
\bP{ K_1(\theta) \cap K_3(\theta)  
=  \emptyset , K_2(\theta) \cap K_3(\theta) = \emptyset }
=
q(\theta)^2 \label{eq:q^2_theta}
\]
and
\begin{eqnarray}
\lefteqn{ 
\bP{ 
\begin{array}{c}
K_1(\theta) \cap K_2(\theta) = \emptyset , K_1(\theta) \cap K_3(\theta) = \emptyset \\
K_2(\theta) \cap K_3(\theta)  = \emptyset  \\
\end{array}
}
} &&
\nonumber \\
&=& \bP{ 
\begin{array}{c}
K_1(\theta) \cap K_2(\theta) = \emptyset , \\
K_3(\theta) \cap ( K_1(\theta) \cup K_2(\theta) ) = \emptyset \\
\end{array}
}
\nonumber \\
&=& q (\theta) r(\theta) 
\nonumber
\end{eqnarray}
since $|K_1(\theta) \cup K_2(\theta)| = 2K$ when $K_1(\theta) \cap K_2(\theta) = \emptyset$.
Collecting these facts we find
\begin{eqnarray}
\lefteqn{
\bE{ \chi_{{123}} (\theta) } 
} & &
\nonumber \\
&=& 
\left ( 1 - q(\theta) \right )^2
- \left ( 1 - q(\theta) \right ) q(\theta) + q(\theta)^2 
- q (\theta) r(\theta)
\nonumber
\end{eqnarray}
and the conclusion (\ref{eq:ProbTriangle}) follows by elementary
algebra.
\myendpf

\subsection{A proof of Proposition \ref{prop:SecondMoment}}
\label{subsec:ProofRKG+SecondMoment}

By exchangeability and the
binary nature of the rvs involved we readily obtain
\begin{eqnarray}
\lefteqn{\bE{ T_n (\theta)^2 }} &&
\nonumber \\ 
&=& \bE{ T_n (\theta)}
\label{eq:SecondMomentPart0} + {n \choose 3}{3 \choose
2}{n-3\choose 1}
        \bE{\chi_{123}(\theta) \chi_{ 124} (\theta)}
\nonumber \\
&&+ {n \choose 3}{3 \choose 1}{n-3\choose 2}
       \bE{\chi_{123}(\theta) \chi_{ 145} (\theta)}
       \nonumber \\
&& + {n \choose 3}{n-3\choose 3}
        \bE{\chi_{123}(\theta) \chi_{ 456} (\theta)}.
\label{eq:SecondMomentInitialFormula}
\end{eqnarray}

Under the enforced independence assumptions the rvs
$\chi_{123}(\theta)$ and $\chi_{ 456} (\theta)$ are
independent and identically distributed. As a result,
\[
\bE{\chi_{123}(\theta) \chi_{ 456} (\theta)} =
\bE{\chi_{123}(\theta)} \bE{ \chi_{ 456} (\theta)} = \beta
(\theta)^2,
\]
and using the relation (\ref{eq:RKG+FirstMoment}) yields
\begin{eqnarray}
{n \choose 3}{n-3\choose 3} 
\bE{\chi_{123}(\theta) \chi_{ 456} (\theta)} 
=
\frac{ {n-3\choose 3} }{ {n \choose 3} } 
 \left ( \bE{ T_n (\theta) } \right )^2.
 \label{eq:SecondMomentPart1} 
\end{eqnarray}

On the other hand, with the help of
(\ref{eq:Probab_key_ring_does_not_intersect_S}) 
we readily check that the
indicator rvs $\chi_{123}(\theta)$ and $\chi_{145}(\theta)$
are independent and identically distributed {\em conditionally} 
on $K_1(\theta)$ with
\begin{equation}
\bP{ \chi_{123}(\theta) = 1 | K_1 (\theta ) } 
= \bP{ \chi_{123}(\theta) = 1 } 
= \beta (\theta) . 
\label{eq:chi_123}
\end{equation}
As a similar statement applies to $\chi_{145}(\theta)$, we
conclude that the rvs $\chi_{123}(\theta)$ and
$\chi_{145}(\theta)$ are (unconditionally) independent and
identically distributed with
\[
\bE{\chi_{123}(\theta) \chi_{ 145} (\theta)} =
\bE{\chi_{123}(\theta) } \bE{ \chi_{ 145} (\theta)} =
\beta(\theta)^2.
\]
Again by virtue of (\ref{eq:RKG+FirstMoment}), this last observation
yields
\begin{eqnarray}
\lefteqn{{n \choose 3}{3 \choose 1}{n-3\choose 2} \bE{\chi_{123}(\theta)
\chi_{ 145} (\theta)}}
\nonumber \\
 &=& 3 \frac{ {n-3\choose 2} }{ {n \choose 3}
} \cdot \left ( \bE{ T_n (\theta) } \right )^2 .
\label{eq:SecondMomentPart2}
\end{eqnarray}
Substituting (\ref{eq:SecondMomentPart1}) and
(\ref{eq:SecondMomentPart2}) into 
(\ref{eq:SecondMomentInitialFormula})
establishes Proposition \ref{prop:SecondMoment}.
\myendpf

\subsection{A proof of Proposition \ref{prop:AsymptoticEquivalence2}}
\label{sec:ProofOfAsymptEquiv2}
Since $1 \leq K_n \leq {K_n}^2$ for all $n=1,2, \ldots $, 
the condition (\ref{eq:Condition1+2}) implies both
\begin{equation}
\lim_{n \rightarrow \infty } \frac{1}{P_n} = 0
\quad \mbox{\rm and} \quad
\lim_{n \rightarrow \infty } \frac{K_n}{P_n} = 0.
\label{eq:RatioConditionStrong+Consequence1}
\end{equation}
Therefore, $\lim_{n \rightarrow \infty } P_n = \infty$, and for any $c > 0$, 
we have $ c K_n < P_n $ for all $n$ sufficiently large in
$\mathbb{N}_0$ (dependent on $c$).
%by the comments
%following (\ref{eq:RatioConditionStrong+Consequence1}),
Thus, we have $3K_n < P_n$ for all $n$ sufficiently large in $\mathbb{N}_0$.
On that range we can use the expression (\ref{eq:beta(theta)}) to write
\[
\beta ( \theta_n ) = \left ( 1 - q(\theta_n) \right )^3
                        +  q ( \theta_n )^3
\left( 1-\frac { r ( \theta_n ) } { q (\theta_n)^2 } \right ).
\]
As Lemma \ref{lem:AsymptoticEquivalence1} already implies 
$q (\theta_n )^ 3 \sim 1$ and $\left ( 1 - q(\theta_n) \right )^3 \sim
\left ( \frac{K^2_n}{P_n} \right )^3 $, the asymptotic equivalence
$\beta(\theta_n) \sim \tau(\theta_n)$ will be established if we
show that
\begin{equation}
1 - \frac{r(\theta_n)}{q(\theta_n)^2} \sim \frac{K^3_n}{P^2_n} .
\label{eq:AsymptoticsEquivalence2Reduced}
\end{equation}
This is an easy consequence of the fact that all terms involved are non-negative.

To establish (\ref{eq:AsymptoticsEquivalence2Reduced}) we proceed as follows:
With positive integers $K, P$  such that $3K \leq P$, we note that
\begin{eqnarray}
& &
\frac{r(\theta)}{q(\theta)^2} 
\nonumber \\
&=& 
\left( \frac { ( P - 2 K ) ! } {
( P - K ) ! } \right ) ^ 2 \cdot \frac { ( P - 2 K ) ! } { ( P - 3 K ) !  } \cdot \frac { P !  } { ( P - K ) ! }
\nonumber \\
&=&
\frac { ( P - 2 K ) ! ( P- 2K)! } { ( P - K ) !  (P-3K)!}  \cdot \frac { P! ( P - 2 K ) ! } { ( P - K ) !  (P-K)!}
\nonumber \\
&=& 
\prod_{\ell=0}^{K-1} \left ( \frac{P-2K-\ell}{P-K-\ell} \right)
\cdot \prod_{\ell=0}^{K-1} \left ( \frac{P-\ell}{P-K-\ell} \right )
\nonumber \\
&=& \prod_{\ell=0}^{K-1} \left ( 1 - \frac{K}{P-K-\ell} \right)
\cdot \prod_{\ell=0}^{K-1} \left ( 1 + \frac{K}{P-K-\ell} \right )
\nonumber \\
&=& \prod_{\ell=0}^{K-1} \left( 1 - \left(\frac{K}{P-K-\ell}\right)^2 \right) 
\nonumber
\end{eqnarray}
upon grouping  factors appropriately. Elementary bounding arguments now yield the two bounds
\[
1- \left(1-\left(\frac{K}{P-K}\right)^2\right)^K \leq 1 - \frac{ r( \theta ) } { q ( \theta ) ^ 2 }
\]
and
\[
1 - \frac{ r( \theta ) } { q ( \theta ) ^ 2 }
\leq 
1 -\left(1-\left(\frac{K}{P-2K}\right)^2\right)^K .
\]

Pick a scaling $P,K: \mathbb{N}_0 \rightarrow \mathbb{N}_0$
satisfying the equivalent conditions (\ref{eq:Condition1+2}) and
consider $n$ sufficiently large in $\mathbb{N}_0$ so that $3K_n <
P_n$. On that range, we replace $\theta $ by $\theta_n $ in the
last chain of inequalities according to this scaling. A standard
sandwich argument will yield the desired equivalence
(\ref{eq:AsymptoticsEquivalence2Reduced}) if we show that
\begin{equation}
1 - \left( 1-\left( \frac{K_n}{P_n-cK_n} \right)^2 \right)^{K_n}
\sim \ \frac{K_n^3}{P^2_n}, \quad c=1,2 .
\label{eq:AsymptoticEquivToBeShown}
\end{equation}

To do so we proceed as follows: Fix $c=1,2$. With
\[
A_n(c) = \left ( \frac{K_n}{P_n-cK_n} \right ), \quad n=1,2,
\ldots
\]
standard calculus yields
\begin{eqnarray}\nonumber
\lefteqn{
1 - \left( 1-\left( \frac{K_n}{P_n-cK_n} \right)^2 \right)^{K_n}
} && 
\nonumber \\
 &=&
 1 - \left( 1- A_n(c)^2 \right )^{K_n}
\nonumber \\
 &=&
K_n A_n(c)^2 \int_0^1 \left ( 1 - A_n(c)^2 t \right )^{K_n -1} dt
\end{eqnarray}
on the appropriate range. The asymptotic equivalences
\begin{equation}
A_n(c)^2 = \left(\frac{K_n}{P_n-cK_n}\right)^2 \sim \left (
\frac{K_n }{P_n} \right )^2 
\label{eq:BasicAsymptoticEquality2}
\end{equation}
and
\begin{equation}
K_n A_n(c)^2
\sim \frac{K^3_n }{P^2_n}
\label{eq:BasicAsymptoticEquality2bb}
\end{equation}
follow from (\ref{eq:RatioConditionStrong+Consequence1}), so that
(\ref{eq:AsymptoticEquivToBeShown}) will hold if we show that
\begin{equation}
\lim_{n \rightarrow \infty} 
\int_{0}^1 \left ( 1 - A_n(c)^2 t \right )^{K_n -1} dt = 1 . 
\label{eq:AsymptoticEquivToBeShown2}
\end{equation}
In view of (\ref{eq:BasicAsymptoticEquality2}) we
conclude from (\ref{eq:RatioConditionStrong+Consequence1}) that
for all $n$ sufficiently large in $\mathbb{N}_0$ we have $
\sup_{0\leq t \leq 1} \left | 1 - A_n(c)^2 t \right | \leq 1$.
Therefore, the Bounded Convergence Theorem will yield
(\ref{eq:AsymptoticEquivToBeShown2}) as soon as we establish
\begin{equation}
\lim_{n \rightarrow \infty} \left ( 1 - A_n(c)^2 t \right )^{K_n
-1} = 1 , \quad 0 \leq t \leq 1 .
\label{eq:AsymptoticEquivToBeShown3}
\end{equation}

To that end, recall the decomposition
\begin{equation}
\log ( 1 - x ) = - \int_0^x \frac{1}{1-t} dt = - x - \Psi (x)
\label{eq:LimitDecompositionBasic}
\end{equation}
where
\[
\Psi (x) = \int_0^x \frac{t}{1-t} dt,
\quad 0 \leq x < 1 .
\]
It is easy to check that
\begin{equation}
\lim_{x \downarrow 0} \frac{ \Psi(x) } {x} = 0.
\label{eq:LimitDecomposition}
\end{equation}

Fix $n$ sufficiently large in $\mathbb{N}_0$ as required above.
For each $t$ in the interval $(0,1]$, with the help of
(\ref{eq:LimitDecompositionBasic}) we can write
\begin{eqnarray}
\lefteqn{\left ( 1 - A_n(c)^2 t \right )^{K_n -1}} &&
\nonumber \\
&=& e^{ (K_n-1) \log \left( 1 - A_n(c)^2 t \right )} 
\nonumber \\
&=& e^{ - (K_n-1) A_n(c)^2 t - (K_n-1)
\Psi ( A_n(c)^2 t ) } . \label{eq:DecompositionExponent}
\end{eqnarray}
Returning to  
(\ref{eq:BasicAsymptoticEquality2bb}), we use
(\ref{eq:Condition1+2}) and
(\ref{eq:RatioConditionStrong+Consequence1}) to find
\[
\lim_{n \to \infty} K_n A_n(c)^2 = \lim_{n \to \infty} \left (
\frac{K^2_n }{P_n} \cdot \frac{K_n}{P_n} \right ) = 0 .
\]
It is then plain that
$ \lim_{n \to \infty} (K_n-1) A_n(c)^2 = 0$, whence
\begin{eqnarray}
\lefteqn{\lim_{n \to \infty} (K_n-1) \Psi ( A_n(c)^2 t )}
\nonumber \\
&=& \lim_{n \to
\infty} (K_n-1) A_n(c)^2 t \cdot \frac{ \Psi ( A_n(c)^2 t ) }{
A_n(c)^2 t } = 0
\nonumber
\end{eqnarray}
with the help of (\ref{eq:LimitDecomposition}) in the last step.
Finally, letting $n$ go to infinity in
(\ref{eq:DecompositionExponent}), we readily get
(\ref{eq:AsymptoticEquivToBeShown3}) as desired.
\myendpf

\section{Proofs of the main results}
\label{sec:proofs}

\subsection{A proof of Theorem \ref{thm:ZeroLaw}}
\label{subsec:ProofZeroLaw}

Consider a scaling $P,K: \mathbb{N}_0\rightarrow\mathbb{N}_0$. 
For each $n=3,4, \ldots $, 
the elementary bound $\bP{ T_n (\theta_n) > 0 } \leq \bE{ T_n(\theta_n) } $ 
implies
\[
\bP{ T_n ( \theta_n) > 0 } \leq {n \choose 3} \beta (\theta_n)
\]
by virtue of Proposition \ref{prop:RKG+FirstMoment}. Theorem
\ref{thm:ZeroLaw} thus follows if under
(\ref{eq:ConditionForZero}) we show that $ \lim_{n \rightarrow
\infty} {n \choose 3} \beta (\theta_n) = 0$. By Proposition
\ref{prop:AsymptoticEquivalence2} this convergence is
equivalent to the assumed condition $\lim_{n \rightarrow \infty}
n^3 \tau(\theta_n) = 0$, and the proof of Theorem
\ref{thm:ZeroLaw} is now complete.
\myendpf

\subsection{A proof of Theorem \ref{thm:OneLaw}}
\label{subsec:ProofOneLaw}

Assume first that $q^\star$ satisfies $0 \leq q^\star < 1$. Fix
$n=3,4, \ldots $ and partition the $n$ nodes into the $k_n+1$
non-overlapping groups $(1,2,3)$, $(4,5,6)$, $\ldots $,
$(3k_n+1,3k_n+2,3k_n+3)$ with $k_n = \lfloor \frac{n-3}{3} \rfloor
$. If $\mathbb{K}(n;\theta_n)$ contains no triangle, then {\em
none} of these $k_n + 1$ groups of nodes forms a triangle. With
this in mind we get
\begin{eqnarray}
\lefteqn{\bP{ T_n(\theta_n) = 0 } } & &
\nonumber \\
&\leq& \bP{ \bigcap_{\ell=0}^{k_n} \left [
\begin{array}{c}
\mbox{Nodes $3\ell+1,3\ell+2, 3\ell+3$ do not} \\
\mbox{form a triangle in $\mathbb{K}(n;\theta_n)$ } \\
\end{array}
\right ] }
\nonumber \\
&=& \prod_{\ell=0}^{k_n} \bP{
\begin{array}{c}
\mbox{Nodes $3\ell+1,3\ell+2, 3\ell+3$ do not} \\
\mbox{form a triangle in $\mathbb{K}(n;\theta_n)$ } \\
\end{array}
}
\label{eq:IndependenceTriangle} \\
&=& \left ( 1 - \beta(\theta_n) \right )^{k_n+1}
\nonumber \\
&\leq& \left ( 1 - (1-q(\theta_n) )^3 \right )^{k_n+1}
\label{eq:OneLawInequality0}
\\
&\leq& e^{- (k_n +1 ) (1-q(\theta_n) )^3 }.
\label{eq:OneLawInequality1}
\end{eqnarray}
Note that (\ref{eq:IndependenceTriangle}) follows from the fact
that the events
\[
\left [
\begin{array}{c}
\mbox{Nodes $3\ell+1,3\ell+2, 3\ell+3$ do not} \\
\mbox{form a triangle in $\mathbb{K}(n;\theta_n)$ } \\
\end{array}
\right ], \; \ell =0, \ldots , k_n
\]
are mutually independent due to the non-overlap condition, while
the inequality (\ref{eq:OneLawInequality0}) is justified with the
help of (\ref{eq:r(tetha)C}). Let $n$ go to infinity in the
inequality (\ref{eq:OneLawInequality1}). From the constraint
$q^\star < 1$ we conclude that 
$\lim_{n \rightarrow \infty} \bP{ T_n(\theta_n )=0 } = 0$ 
since $k_n \sim \frac{n}{3}$ so that
$\lim_{n \rightarrow \infty} ( k_n + 1 ) (1-q(\theta_n) )^3 =
\infty $. This establishes the one law in the case $q^\star < 1$.
 
To handle the case $q^\star =1$,
we use a standard bound which forms the basis of the method of
second moment \cite[Remark 3.1, p. 55]{JansonLuczakRucinski}. Here
this bound takes the form
\begin{equation}
\frac{ \left ( \bE{ T_n (\theta_n)} \right )^2}
     { \bE{T_n (\theta_n)^2 }} \leq
\bP{T_n (\theta_n) > 0 }, \quad n=3,4, \ldots
\label{eq:SecondMoment+b}
\end{equation}
Theorem \ref{thm:OneLaw} then 
will be established in the case $q^\star =1 $ 
if we show under (\ref{eq:Condition1+2}) that the condition
(\ref{eq:ConditionForOne}) implies
\begin{equation}
\lim_{n \rightarrow \infty } \frac{\bE{T_n (\theta_n)^2}}
     {\left ( \bE{T_n (\theta_n)} \right )^2} = 1.
\label{eq:OneLawConvergenceSecondMoment}
\end{equation}

As pointed earlier, the conditions (\ref{eq:Condition1+2}) imply $3 K_n < P_n$
for all $n$ sufficiently large in $\mathbb{N}_0$. On that range,
with $\theta$ replaced by $\theta_n$, Proposition
\ref{prop:SecondMoment} yields
\begin{eqnarray}
\frac{ \bE{ T_n (\theta_n)^2 } }
     { \left ( \bE{ T_n (\theta_n) } \right )^2 }
&=& \frac{1}{ \bE{ T_n (\theta_n) } } 
+ \left ( 
\frac{ {n-3\choose 3} }{ {n \choose 3} } 
+ 3 \frac{ {n-3\choose 2} }{ {n \choose 3} }
\right ) 
\nonumber \\ 
& & ~ + \frac{ 3(n-3) }{ {n \choose 3} } 
\cdot 
\frac{ \bE{\chi_{123}(\theta_n) \chi_{ 124} (\theta_n)} }
     { \left ( \bE{\chi_{123}(\theta_n) } \right )^2 }
\nonumber
\end{eqnarray}
as we make use of (\ref{eq:RKG+FirstMoment})
in the last term.

Let $n$ go to infinity in the resulting expression: 
Under condition (\ref{eq:ConditionForOne}), 
we have 
$\lim_{n\rightarrow \infty} n^3 \beta (\theta_n) = \infty$ 
by Proposition \ref{prop:AsymptoticEquivalence2},
whence 
$\lim_{n\rightarrow \infty} \bE{ T_n(\theta_n) } = \infty$ 
by virtue of (\ref{eq:RKG+FirstMoment}).
Since
\begin{equation}
\lim_{n \rightarrow \infty} 
\left ( 
\frac{ {n-3\choose 3} }{ {n \choose 3} } 
+ 3 \frac{ {n-3\choose 2} }{ {n \choose 3} } 
\right )
= 1 
\label{eq:LimitCoefficient1}
\end{equation} 
and
\begin{equation}
\frac{ {n \choose 3} }{ 3(n-3) } \sim \frac{n^2}{18},
\label{eq:LimitCoefficient2}
\end{equation}
the convergence (\ref{eq:OneLawConvergenceSecondMoment}) 
will hold if we show that
\begin{equation}
\lim_{n \rightarrow \infty } \frac{1}{n^2} \frac{
\bE{\chi_{123}(\theta_n) \chi_{ 124} (\theta_n)} }
     { \left ( \bE{\chi_{123}(\theta_n) } \right )^2 }
= 0 \label{eq:ToBeShown}
\end{equation}
under the foregoing conditions on the scaling.

This is shown as follows: 
Given positive integers $K$ and $P$ such
that $K \leq P$, fix $n=3, 4, \ldots $. It is immediate that
\begin{eqnarray}
\lefteqn{\bE{\chi_{123}(\theta) \chi_{ 124} (\theta)}}
\nonumber \\ 
&\leq& \bE{
\chi_{123}(\theta)\1{K_1(\theta)\cap K_4(\theta) \neq \emptyset}
}. \label{eq:key_ineq}
\end{eqnarray}
From (\ref{eq:Probab_key_ring_does_not_intersect_S}) it follows
that the rvs $\chi_{123}(\theta)$ and $\1{K_1(\theta)\cap
K_4(\theta)\neq \emptyset}$ are independent conditionally on $K_1
(\theta)$, and an easy conditioning argument yields
\begin{equation}
\bE{\chi_{123}(\theta)
    \1{K_1(\theta)\cap K_4(\theta)\neq \emptyset}}
= \beta(\theta) (1-q(\theta)) \label{eq:key_indp}
\end{equation}
as we recall (\ref{eq:q_theta}) and (\ref{eq:ProbTriangle}). Using
(\ref{eq:key_ineq}) together with (\ref{eq:ProbTriangle}) and
(\ref{eq:key_indp}) we readily obtain the inequalities
\begin{equation}
\frac{ \bE{\chi_{123}(\theta) \chi_{ 124} (\theta)} }
     { \left ( \bE{\chi_{123}(\theta) } \right )^2 }
\leq \frac{ \beta(\theta) (1-q(\theta)) }
     { \beta (\theta)^2 }
\leq \beta(\theta)^{-2/3} 
\label{eq:key_combined}
\end{equation}
where in the last step we noted that 
$1-q(\theta) \leq \beta(\theta)^{1/3}$ 
by appealing to (\ref{eq:r(tetha)C}).

Returning to the convergence (\ref{eq:ToBeShown}) we
see from (\ref{eq:key_combined}) that we need only show
\begin{equation}
\lim_{n \rightarrow \infty} n^2 \beta(\theta_n)^{2/3} = \infty .
\label{eq:toshow2}
\end{equation}
As Proposition \ref{prop:AsymptoticEquivalence2} yields $ n^2
\beta(\theta_n)^{2/3} \sim n^2 \tau(\theta_n)^{2/3} = \left(n^3
\tau(\theta_n)\right)^{2/3}, $ the desired conclusion
(\ref{eq:toshow2}) follows under the condition
(\ref{eq:ConditionForOne}).
\myendpf

\subsection{A proof of Theorem \ref{thm:Equiv_of_clust_defn_RKG}}
\label{sec:RKGConvergenceClustering}

Throughout $P$ and $K$ are positive integers such that
$K \leq P$, and fix $n=3,4, \ldots $.
For each $i=1, \ldots , n$, we introduce the index set 
\begin{equation}
{\cal P}_{n,i}
= \left \{
(j,k): \ 1 \leq j < k \leq n , \ j \neq i, \ k \neq i 
\right \} .
\label{eq:Pairs}
\end{equation}
Next, define the count rvs
$T_{n,i}(\theta)$ and $T^\star_{n,i}(\theta)$
by
\[
T_{n,i}(\theta) 
= 
\sum_{ (j,k) \in {\cal P}_{n,i} }
\xi_{ij}(\theta) \xi_{ik}(\theta) \xi_{jk}(\theta)
\]
and
\[
T^\star_{n,i}(\theta) 
= 
\sum_{ (j,k) \in {\cal P}_{n,i} } \xi_{ij}(\theta) \xi_{ik}(\theta) .
\]
The rv $T_{n,i}(\theta)$ counts the number of distinct triangles
in $\mathbb{K}(n;\theta)$ which have node $i$ as a vertex, while
$T^\star_{n,i}(\theta)$ counts the number of (unordered) distinct pairs of
nodes which are both connected to node $i$ in $\mathbb{K}(n;\theta)$.
The rv $D_{n,i}(\theta)$ is the degree of node $i$ in
$\mathbb{K}(n;\theta)$ and is given by
\[
D_{n,i}(\theta)
=
\sum_{k=1, \ k \neq i }^n \xi_{ik}(\theta).
\]

We have
\begin{align}
\sum_{i=1}^n T_{n,i}(\theta)
= 3 T_n(\theta) 
\nonumber
\end{align}
while
\begin{align}
D_{n,i}(\theta) \left ( D_{n,i}(\theta) - 1 \right )
=
2 T^\star_{n,i}(\theta).
\nonumber
\end{align}
Under the condition
\[
\sum_{i=1}^n D_{n,i}(\theta) \left ( D_{n,i}(\theta) - 1 \right ) > 0,
\]
the definition of $C^\star( \mathbb{K}(n;\theta) )$ yields
\begin{eqnarray}
C^\star( \mathbb{K}(n;\theta) )
&=& 
\frac{ \sum_{i=1}^n T_{n,i}(\theta) }
        { \frac{1}{2} \sum_{i=1}^n D_{n,i}(\theta) \left ( D_{n,i}(\theta) - 1 \right )
 }
\nonumber \\
&=&
\frac{ \sum_{i=1}^n T_{n,i}(\theta) }
        { \sum_{i=1}^n T^\star_{n,i}(\theta) }
\nonumber 
\end{eqnarray}
so that
\begin{equation}
C^\star( \mathbb{K}(n;\theta) )
= 
\frac{ 3 T_n(\theta) }
     { \sum_{i=1}^n T^\star_{n,i}(\theta) }
\1{ \sum_{i=1}^n T^\star_{n,i}(\theta) > 0 } .
\label{eq:ClusteringCoefficientRatio_n}
\end{equation}
The desired  conclusion
(\ref{eq:Equiv_of_clust_defn_RKG}) is now immediate from
Lemma \ref{lem:TopConvergence}
and Lemma \ref{lem:BottomConvergence} established below.
They deal with the a.s. convergence of the numerator
and denominator (properly normalized) appearing in the ratio
(\ref{eq:ClusteringCoefficientRatio_n}), respectively.

\begin{lemma}
{\sl 
For positive integers $P$ and $K$ such that $K \leq P$, we have
\begin{equation}
\lim_{n \rightarrow \infty}
\frac{ T_n (\theta) }{ {n \choose 3} }
= \beta (\theta) 
\quad a.s.
\label{eq:TopConvergence}
\end{equation}
}
\label{lem:TopConvergence}
\end{lemma}

\myproof
Fix $n=3,4, \ldots $ and $\varepsilon > 0$.
Markov's inequality already gives
\[
\bP{ 
\left | \frac{ T_n (\theta) }{ {n \choose 3} }  - \beta (\theta ) \right |
> \varepsilon 
}
\leq \varepsilon^{-2}  {\rm Var} \left [ \frac{ T_n (\theta) }{ {n \choose 3} } \right ]
\]
as we recall (\ref{eq:RKG+FirstMoment}).
It is now plain from (\ref{eq:SecondMoment}) that
\begin{align}
& {\rm Var} \left [ \frac{ T_n (\theta) }{ {n \choose 3} } \right ]
\nonumber \\
& ~ =
\bE{ \left ( \frac{ T_n (\theta) } { {n \choose 3} } \right )^2 }
-
\left ( \frac{ \bE{ T_n (\theta) } } { {n \choose 3} } \right )^2 
\nonumber \\
& ~ =
\frac{ \bE{ T_n (\theta) } }{ {n \choose 3} ^2 }
+ \left ( \frac{
{n-3\choose 3} }{ {n \choose 3} } + 3 \frac{ {n-3\choose 2} }{ {n
\choose 3} }  - 1 \right ) 
\cdot 
\left ( \frac{ \bE{ T_n (\theta) } } { {n \choose 3} } \right )^2 
\nonumber \\
&  ~~~~~ + 3 (n-3) {n \choose 3}
 \cdot
\frac{ \bE{\chi_{123}(\theta) \chi_{ 124} (\theta)} }
     { {n \choose 3}^2 }
\nonumber \\
& ~ =
\frac{ \beta (\theta) }{ {n \choose 3} }
+ \left ( \frac{
{n-3\choose 3} }{ {n \choose 3} } + 3 \frac{ {n-3\choose 2} }{ {n
\choose 3} }  - 1 \right )
\cdot \beta (\theta)^2 
\nonumber \\
&  ~~~~~~ + \frac{ 3(n-3) }{ {n \choose 3} }  \cdot
\bE{ \chi_{123}(\theta) \chi_{ 124} (\theta)}
\label{eq:ExpressionForNormalizedVariance}
\end{align}
as we again make use of the expression (\ref{eq:RKG+FirstMoment}).

With the help of
(\ref{eq:LimitCoefficient1}) and (\ref{eq:LimitCoefficient2}),
it is easy to see that 
\begin{equation}
\lim_{n \rightarrow \infty} 
{\rm Var} \left [ \frac{ T_n (\theta) }{ {n \choose 3} } \right ]
= 0,
\label{eq:NormalizedVarianceVanishes}
\end{equation}
a fact which would readily imply a weaker form of (\ref{eq:TopConvergence})
with a.s. convergence replaced by convergence in probability.
However, elementary algebra on (\ref{eq:ExpressionForNormalizedVariance})
shows that (\ref{eq:NormalizedVarianceVanishes}) takes place according to
\[
\lim_{n \rightarrow \infty} 
n^2 {\rm Var} \left [ \frac{ T_n (\theta) }{ {n \choose 3} } \right ]
=  C 
\]
with
\[
C =18 \left (  \bE{ \chi_{123}(\theta) \chi_{ 124} (\theta)}  - \beta(\theta)^2 \right ) > 0.
\]
As a result, for every $\varepsilon > 0$, we have
\[
\sum_{n=3}^\infty 
\bP{ 
\left | \frac{ T_n (\theta) }{ {n \choose 3} }  - \beta (\theta ) \right | > \varepsilon 
} \leq 
\frac{C^\prime }{ \varepsilon^2 } \sum_{n=3}^\infty n^{-2} < \infty
\]
for some $C^\prime > C$,
and the conclusion (\ref{eq:TopConvergence}) follows by the Borel-Cantelli Lemma.
\myendpf

\begin{lemma}
{\sl 
For positive integers $P$ and $K$ such that $K \leq P$, we have
\begin{equation}
\lim_{n \rightarrow \infty}
\frac{ \sum_{i=1}^n T^\star_{n,i} (\theta) }{ { n \choose 3} } 
= 3p(\theta)^2 
\quad a.s.
\label{eq:BottomConvergence}
\end{equation}
}
\label{lem:BottomConvergence}
\end{lemma}

\myproof
Fix $n=3,4, \ldots $.
Note that
\begin{eqnarray}
T^\star_{n,1}(\theta)
&=&
\sum_{j=2}^{n-1} \sum_{k=j+1}^n \xi_{1j}(\theta)\xi_{1k}(\theta)
\nonumber \\
&=&
\Phi_n ( \xi_{12} (\theta) , \ldots , \xi_{1n}(\theta) )
\end{eqnarray}
where the mapping 
$\Phi_n : [0,1]^{n-1} \rightarrow \mathbb{R}_+$ is given by
\begin{eqnarray}
\Phi_n ( x_2, \ldots , x_n )
&=& \sum_{\ell=2}^{n-1} \sum_{k=\ell+1}^n x_\ell x_k
\nonumber \\
&=& \sum_{\ell=2}^{n-1} x_\ell \left ( \sum_{k=\ell+1}^n x_k \right )
\label{eq:PhiDefn}
\end{eqnarray}
with $(x_2, \ldots , x_n)$ arbitrary in $[0,1]^{n-1}$.
For each $j=2, \ldots , n$, consider
pairs of  elements $(x_2, \ldots , x_n )$ and $(y_2, \ldots , y_n )$ 
in $[0,1]^{n-1}$  which differ only in the $j^{th}$ component, i.e.,
\[
x_\ell = y_\ell , 
\quad 
%\begin{array}{c}
\ell \neq j, \quad
\ell =2, \ldots , n.
\]
Under such conditions, it is easy to check that
\begin{eqnarray}
\lefteqn{
\left |
\Phi_n ( x_2, \ldots , x_n ) - \Phi_n ( y_2, \ldots , y_n )
\right |
} & &
\nonumber \\
&\leq&
\left | x_j - y_j \right | \cdot \sum_{\ell=2, \ \ell \neq j}^{n-1} x_\ell
\nonumber \\
&\leq&
n-1.
\label{eq:McDiarmidConditions}
\end{eqnarray}

Recall that
the $(n-1)$ rvs $\{ \xi_{1j}(\theta), \ j=2, \ldots , n \}$
are i.i.d. Bernoulli rvs.
In view of the constraints (\ref{eq:McDiarmidConditions})
we can now apply McDiarmid's inequality
\cite{McDiarmid1989} (with $c_j = (n-1)$ for all $j=2, \ldots , n-1$); see
also Corollary 2.17 and Remark 2.28 in the monograph
\cite[p. 38]{JansonLuczakRucinski}.
Thus, for every $t > 0$ we find
\begin{equation}
\bP{ 
\left | T^\star_{n,1}(\theta) - \bE{ T^\star_{n,1}(\theta) } \right | > t }
\leq
2 e^{ - \frac{ 2 t^2 }{ (n-1)^3 } } 
\label{eq:McDiarmidBasic}
\end{equation}
with
\begin{eqnarray}
\bE{ T^\star_{n,1}(\theta) }
&=&
\sum_{j=2}^{n-1} \sum_{k=j+1}^n \bE{ \xi_{1j}(\theta)\xi_{1k}(\theta) }
\nonumber \\
&=& \sum_{j=2}^{n-1} (n-j) p(\theta)^2
\nonumber \\
&=& \frac{ (n-1)(n-2) }{2} \cdot p(\theta)^2
\label{eq:McDiarmidMoment}
\end{eqnarray}
under the independence noted earlier.

With $\varepsilon > 0$ we now substitute
\begin{equation}
t = \frac{(n-1)(n-2)}{2} \varepsilon
\label{eq:t}
\end{equation}
into (\ref{eq:McDiarmidBasic}). Since
\begin{equation}
\frac{2t^2}{(n-1)^3 }
= \frac{ (n-2)^2 }{ 2 (n-1) } \cdot \varepsilon^2
\sim \frac{ n }{2 } \cdot \varepsilon^2,
\label{eq:Ratio}
\end{equation}
we obtain from (\ref{eq:McDiarmidBasic}) and (\ref{eq:McDiarmidMoment}) that
\begin{equation}
\bP{ 
\left | 
\frac{ T^\star_{n,1}(\theta)} { \frac{ (n-1)(n-2) }{2} }  - p(\theta)^2 
\right |
> \varepsilon  }
\leq
2 e^{ - \frac{n}{2} (1 + o(1) ) \varepsilon ^2 }.
\label{eq:McDiarmidExponentialDecay}
\end{equation}
Since
\begin{align}
\left |
\frac{ \sum_{i=1}^n T^\star_{n,i}(\theta) }
     { \frac{n(n-1)(n-2)}{2} }
- p(\theta)^2 
\right |
&=
\left |
\frac{1}{n}
\sum_{i=1}^n 
\left (
\frac{ T^\star_{n,i} (\theta) } { \frac{(n-1)(n-2)}{2} }
- p(\theta)^2 
\right )
\right |
\nonumber \\
&\leq
\frac{1}{n}
\sum_{i=1}^n 
\left |
\frac{ T^\star_{n,i}(\theta)  } { \frac{(n-1)(n-2)}{2} }
- p(\theta)^2
\right |,
\nonumber
\end{align}
it is plain that
\begin{eqnarray}
\lefteqn{
\bP{ 
\left | 
\frac{ \sum_{i=1}^n T^\star_{n,i}(\theta) }
     { \frac{n(n-1)(n-2)}{2} }
- p(\theta)^2 
\right |
> \varepsilon
} 
} & & 
\nonumber \\
&\leq&
\bP{
\frac{1}{n}
\sum_{i=1}^n 
\left |
\frac{ T^\star_{n,i}(\theta) } { \frac{(n-1)(n-2)}{2} }
- p(\theta)^2
\right |
> \varepsilon
}
\nonumber \\
&\leq&
\bP{
\bigcup_{i=1}^n 
\left [
\left |
\frac{ T^\star_{n,i}(\theta) } { \frac{(n-1)(n-2)}{2} }
- p(\theta)^2
\right |
> \varepsilon 
\right ]
}
\nonumber \\
&\leq&
\sum_{i=1}^n 
\bP{
\left |
\frac{ T^\star_{n,i} (\theta) } { \frac{(n-1)(n-2)}{2} }
- p(\theta)^2
\right |
> \varepsilon 
}
\nonumber \\
&=&
n 
\bP{
\left |
\frac{ T^\star_{n,1} (\theta) } { \frac{(n-1)(n-2)}{2} }
- p(\theta)^2
\right |
> \varepsilon 
}
\label{eq:RKG+UnionBound}
\end{eqnarray}
where the last inequality follows by a union bound argument
and (\ref{eq:RKG+UnionBound}) is a consequence of exchangeability.

Invoking (\ref{eq:McDiarmidExponentialDecay}) (with $\frac{\varepsilon}{3}$ instead of $\varepsilon$) we get
\[
\bP{ 
\left | 
\frac{ \sum_{i=1}^n T^\star_{n,i}(\theta)  }
     { {n \choose 3} }
- 3 p(\theta)^2 
\right |
> \varepsilon
} 
\leq 
2 n e^{ - \frac{n}{18} (1 + o(1) ) \varepsilon ^2 }
\]
with
\[
\sum_{n=3}^\infty n e^{ - \frac{n}{18} (1 + o(1) ) \varepsilon ^2 }
< \infty
\]
for every $\varepsilon > 0$.
The a.s. convergence (\ref{eq:BottomConvergence}) now follows by 
the Borel-Cantelli Lemma.
\myendpf

\section*{\large Appendix}

\setcounter{equation}{0}
\setcounter{section}{0}
\renewcommand{\thesection}{\Alph{section}}
\renewcommand{\theequation}{\thesection.\arabic{equation}}

\section{A proof of Theorem \ref{thm:Equiv_of_clust_defn_ER}}
\label{sec:ERConvergenceClustering}

The pattern of proof is very similar to that given for 
Theorem \ref{thm:Equiv_of_clust_defn_RKG} in Appendix \ref{sec:RKGConvergenceClustering}:
Throughout pick $p$ in $(0,1)$ and fix $n=3,4, \ldots $.
With distinct nodes $i,j=1, \ldots , n$, introduce the indicator function
\[
\xi_{ij}(p) = \1{ E_{ij}(p) }.
\]

As in the proof of Theorem \ref{thm:Equiv_of_clust_defn_RKG},
for each $i=1, \ldots , n$, we define the rvs $T_{n,i}(p)$ and $T^\star_{n,i}(p)$
by
\[
T_{n,i}(p) 
= 
\sum_{ (j,k) \in {\cal P}_{n,i} }
\xi_{ij}(p) \xi_{ik}(p) \xi_{jk}(p)
\]
and
\[
T^\star_{n,i}(p) 
= 
\sum_{ (j,k) \in {\cal P}_{n,i} } \xi_{ij}(p) \xi_{ik}(p) 
\]
with index set $\mathcal{P}_{n,i}$ defined by (\ref{eq:Pairs}).
The rv $T_{n,i}(p)$ counts the number of distinct triangles
in $\mathbb{G}(n;p)$ which have node $i$ as a vertex, while
$T^\star_{n,i}(p)$ counts the number of (unordered) distinct pairs of
nodes which are both connected to node $i$ in $\mathbb{G}(n;p)$.
The degree $D_{n,i}(p)$ of node $i$ in $\mathbb{G}(n;p)$ is given by
\[
D_{n,i}(p)
=
\sum_{k=1, \ k \neq i }^n \xi_{ik}(p).
\]

Again we have the relations
\begin{align}
\sum_{i=1}^n T_{n,i}(p)
= 3 T_n(p) 
\nonumber
\end{align}
and
\begin{align}
D_{n,i}(p) \left ( D_{n,i}(p) - 1 \right )
=
2 T^\star_{n,i}(p).
\nonumber
\end{align}
Under the condition
\[
\sum_{i=1}^n D_{n,i}(p) \left ( D_{n,i}(p) - 1 \right ) > 0,
\]
the definition of $C^\star( \mathbb{G}(n;p) )$ yields
\begin{eqnarray}
C^\star( \mathbb{G}(n;p) )
&=& 
\frac{ \sum_{i=1}^n T_{n,i}(p) }
        { \frac{1}{2} \sum_{i=1}^n D_{n,i}(p) \left ( D_{n,i}(p) - 1 \right )
 }
\nonumber \\
&=&
\frac{ \sum_{i=1}^n T_{n,i}(p) }
        { \sum_{i=1}^n T^\star_{n,i}(p) }
\nonumber 
\end{eqnarray}
so that
\begin{equation}
C^\star( \mathbb{G}(n;p) )
= 
\frac{ 3 T_n(p) }
     { \sum_{i=1}^n T^\star_{n,i}(p) }
\1{ \sum_{i=1}^n T^\star_{n,i}(p) > 0 } .
\label{eq:ERClusteringCoefficientRatio_n}
\end{equation}
The desired  conclusion
(\ref{eq:Equiv_of_clust_defn_ER}) is now immediate from
Lemma \ref{lem:ERTopConvergence}
and Lemma \ref{lem:ERBottomConvergence} established below.
They deal with the a.s. convergence of the numerator
and denominator (properly normalized) appearing in the ratio
(\ref{eq:ERClusteringCoefficientRatio_n}), respectively.

\begin{lemma}
{\sl 
For every $p$ in $(0,1)$, we have
\begin{equation}
\lim_{n \rightarrow \infty}
\frac{ T_n (p) }{ {n \choose 3} }
= \tau^\star(p)
\quad a.s.
\label{eq:ERTopConvergence}
\end{equation}
}
\label{lem:ERTopConvergence}
\end{lemma}

\myproof
Fix $n=3,4, \ldots $ and $\varepsilon > 0$.
Markov's inequality already gives
\[
\bP{ 
\left | \frac{ T_n (p) }{ {n \choose 3} }  - \tau^\star (p) \right |
> \varepsilon 
}
\leq \varepsilon^{-2}  {\rm Var} \left [ \frac{ T_n (p) }{ {n \choose 3} } \right ]
\]
as we recall (\ref{eq:ER+FirstMoment}).

As in the proof of Proposition \ref{prop:SecondMoment}, we readily obtain
\begin{eqnarray}
\lefteqn{\bE{ T_n (p)^2 }} &&
\nonumber \\ 
&=& \bE{ T_n (p)}
\label{eq:ERSecondMomentPart0} + {n \choose 3}{3 \choose
2}{n-3\choose 1}
        \bE{\chi_{123}(\theta) \chi_{ 124} (p)}
\nonumber \\
&&+ {n \choose 3}{3 \choose 1}{n-3\choose 2}
       \bE{\chi_{123}(p) \chi_{ 145} (p)}
       \nonumber \\
&& + {n \choose 3}{n-3\choose 3}
        \bE{\chi_{123}(p) \chi_{ 456} (p)}.
\label{eq:ERSecondMomentInitialFormula}
\end{eqnarray}
by the exchangeability and binary nature of the rvs involved.
Under the assumed independence, we find
\[
 \bE{\chi_{123}(p) \chi_{ 145} (p)}
 =
\bE{\chi_{123}(p) } \bE{ \chi_{ 145} (p)}
  = p^6
\]
and
\[
\bE{\chi_{123}(p) \chi_{ 456} (p)}
=
\bE{\chi_{123}(p) } \bE{ \chi_{ 456} (p)}
= p^6
\]
together with $\bE{\chi_{123}(p) \chi_{ 124} (p)} = p^5$.

Substituting into (\ref{eq:ERSecondMomentInitialFormula}) gives
\begin{eqnarray}
\bE{ T_n (p)^2 } 
&=& \bE{ T_n (p)} + 3 (n-3)  {n \choose 3}p^5  
\nonumber \\
&&+ {n \choose 3}
\left ( 3 {n-3\choose 2} + {n-3\choose 3} \right ) p^6 .
\nonumber 
%\label{eq:ERSecondMomentInitialFormula2}
\end{eqnarray}

It follows that
\begin{align}
& {\rm Var} \left [ \frac{ T_n (p) }{ {n \choose 3} } \right ]
\nonumber \\
& ~ =
\bE{ \left ( \frac{ T_n (p) } { {n \choose 3} } \right )^2 }
-
\left ( \frac{ \bE{ T_n (p) } } { {n \choose 3} } \right )^2 
\nonumber \\
& ~ =
\frac{ \bE{ T_n (p) } }{ {n \choose 3} ^2 }
+ \left ( \frac{
{n-3\choose 3} }{ {n \choose 3} } + 3 \frac{ {n-3\choose 2} }{ {n \choose 3} }  - 1 \right ) 
\cdot 
\left ( \frac{ \bE{ T_n (p) } } { {n \choose 3} } \right )^2 
\nonumber \\
&  ~~~~~ + 3 (n-3) {n \choose 3} 
 \cdot
\frac{ p^5 } { {n \choose 3}^2 }
\label{eq:ERExpressionForNormalizedVariance}
\end{align}
as we again make use of the expression (\ref{eq:ER+FirstMoment}).

With the help of
(\ref{eq:LimitCoefficient1}) and (\ref{eq:LimitCoefficient2}),
it is easy to see that 
\begin{equation}
\lim_{n \rightarrow \infty} 
{\rm Var} \left [ \frac{ T_n (p) }{ {n \choose 3} } \right ]
= 0.
\label{eq:ERNormalizedVarianceVanishes}
\end{equation}
This would readily imply a weaker form of (\ref{eq:ERTopConvergence})
with a.s. convergence replaced by convergence in probability.
However, elementary algebra on (\ref{eq:ERExpressionForNormalizedVariance})
shows that (\ref{eq:ERNormalizedVarianceVanishes}) takes place according to
\[
\lim_{n \rightarrow \infty} 
n^2 {\rm Var} \left [ \frac{ T_n (p) }{ {n \choose 3} } \right ]
=  C
\]
with 
$C =  18 p^5 (1-p)$.
As a result, for every $\varepsilon > 0$, we have
\[
\sum_{n=3}^\infty 
\bP{ 
\left | \frac{ T_n (p) }{ {n \choose 3} }  - \tau^\star (p) \right | > \varepsilon 
} \leq 
\frac{C^\prime }{ \varepsilon^2 } \sum_{n=3}^\infty n^{-2} < \infty
\]
for some $C^\prime > C$,
and the conclusion (\ref{eq:ERTopConvergence}) follows by the Borel-Cantelli Lemma.
\myendpf

\begin{lemma}
{\sl 
For every $p$ in $(0,1)$, we have
\begin{equation}
\lim_{n \rightarrow \infty}
\frac{ \sum_{i=1}^n T^\star_{n,i} (p) }{ { n \choose 3} } 
= 3p^2 
\quad a.s.
\label{eq:ERBottomConvergence}
\end{equation}
}
\label{lem:ERBottomConvergence}
\end{lemma}

\myproof
Fix $n=3,4, \ldots $ and $p$ in $(0,1)$
Again we have
\begin{eqnarray}
T^\star_{n,1}(p)
&=&
\sum_{j=2}^{n-1} \sum_{k=j+1}^n \xi_{1j}(p)\xi_{1k}(p)
\nonumber \\
&=&
\Phi_n ( \xi_{12} (p) , \ldots , \xi_{1n}(p) )
\end{eqnarray}
where the mapping 
$\Phi_n : [0,1]^{n-1} \rightarrow \mathbb{R}_+$ is given by (\ref{eq:PhiDefn}).

The $(n-1)$ rvs $\{ \xi_{1j}(p), \ j=2, \ldots , n \}$ are i.i.d. Bernoulli rvs.
In view of the constraints (\ref{eq:McDiarmidConditions})
we can now apply McDiarmid's inequality
\cite{McDiarmid1989} (with $c_j = (n-1)$ for all $j=2, \ldots , n-1$); see
also Corollary 2.17 and Remark 2.28 in the monograph
\cite[p. 38]{JansonLuczakRucinski}.
Thus, for every $t > 0$ we find
\begin{equation}
\bP{ 
\left | T^\star_{n,1}(p) - \bE{ T^\star_{n,1}(p) } \right |
> t }
\leq
2 e^{ - \frac{ 2 t^2 }{ (n-1)^3 } } 
\label{eq:ERMcDiarmidBasic}
\end{equation}
with
\begin{eqnarray}
\bE{ T^\star_{n,1}(p) }
&=&
\sum_{j=2}^{n-1} \sum_{k=j+1}^n \bE{ \xi_{1j}(p)\xi_{1k}(p) }
\nonumber \\
&=& \sum_{j=2}^{n-1} (n-j) p^2
\nonumber \\
&=& \frac{ (n-1)(n-2) }{2} \cdot p^2
\label{eq:ERMcDiarmidMoment}
\end{eqnarray}
under the assumed independence assumptions.

With $\varepsilon > 0$ we now substitute $t$ given by (\ref{eq:t}) into (\ref{eq:ERMcDiarmidBasic}).
Using (\ref{eq:Ratio}) we obtain
from (\ref{eq:ERMcDiarmidBasic}) and (\ref{eq:ERMcDiarmidMoment}) that
\begin{equation}
\bP{ 
\left | 
\frac{ T^\star_{n,1}(p)} { \frac{ (n-1)(n-2) }{2} }  - p^2 
\right |
> \varepsilon  }
\leq
2 e^{ - \frac{n}{2} (1 + o(1) ) \varepsilon ^2 }.
\label{eq:ERMcDiarmidExponentialDecay}
\end{equation}

The arguments leading to (\ref{eq:RKG+UnionBound}) 
also yield
\begin{eqnarray}
\lefteqn{
\bP{ 
\left | 
\frac{ \sum_{i=1}^n T^\star_{n,i}(p) }{ \frac{n(n-1)(n-2)}{2} } - p^2 
\right |
> \varepsilon
} 
} & & 
\nonumber \\
&\leq&
n 
\bP{
\left |
\frac{ T^\star_{n,1}(p) } { \frac{(n-1)(n-2)}{2} }
- p^2
\right |
> \varepsilon .
}
\nonumber
\end{eqnarray}

For every $\varepsilon > 0$,
invoking (\ref{eq:ERMcDiarmidExponentialDecay}) (with $\frac{\varepsilon}{3}$ instead of $\varepsilon$) we get
\[
\bP{ 
\left | 
\frac{ \sum_{i=1}^n T^\star_{n,i} (p)}{ {n \choose 3} } - 3 p^2 
\right |
> \varepsilon
} 
\leq 
2 n e^{ - \frac{n}{18} (1 + o(1) ) \varepsilon ^2 }
\]
with
\[
\sum_{n=3}^\infty n e^{ - \frac{n}{18} (1 + o(1) ) \varepsilon ^2 }
< \infty .
\]
The a.s. convergence (\ref{eq:ERBottomConvergence}) now follows by 
the Borel-Cantelli Lemma.
\myendpf

\end{document}